\documentclass[aps,prd,superscriptaddress,showpacs,twocolumn,preprintnumbers,nofootinbib]{revtex4}
\usepackage{graphicx}
\usepackage{dcolumn}
\usepackage{bm}
\usepackage{natbib}
\usepackage{amsmath}
\usepackage{amssymb}
\usepackage{epstopdf}

\newcommand{\be}{\begin{equation}}
\newcommand{\ee}{\end{equation}}
\newcommand{\bea}{\begin{eqnarray}}
\newcommand{\eea}{\end{eqnarray}}

\begin{document}

\title{Indirect Dark Matter Signatures in the Cosmic Dark Ages I.

Generalizing the Bound on $s$-wave Dark Matter Annihilation from \emph{Planck}}

\preprint{
MIT-CTP/4682}

\author{Tracy R. Slatyer}
\email{tslatyer@mit.edu}
\affiliation{Center for Theoretical Physics, Massachusetts Institute of Technology, Cambridge, MA 02139, USA}


\begin{abstract} Recent measurements of the cosmic microwave background (CMB) anisotropies by \emph{Planck} provide a sensitive probe of dark matter annihilation during the cosmic dark ages, and specifically constrain the annihilation parameter $f_\mathrm{eff} \langle \sigma v \rangle/m_\chi$. Using new results (Paper II) for the ionization produced by particles injected at arbitrary energies, we calculate and provide $f_\mathrm{eff}$ values for photons and $e^+e^-$ pairs injected at keV-TeV energies; the $f_\mathrm{eff}$ value for any dark matter model can be obtained straightforwardly by weighting these results by the spectrum of annihilation products. This result allows the sensitive and robust constraints on dark matter annihilation presented by the \emph{Planck} Collaboration to be applied to arbitrary dark matter models with $s$-wave annihilation. We demonstrate the validity of this approach using principal component analysis. As an example, we integrate over the spectrum of annihilation products for a range of Standard Model final states to determine the CMB bounds on these models as a function of dark matter mass, and demonstrate that the new limits generically exclude models proposed to explain the observed high-energy rise in the cosmic ray positron fraction. We make our results publicly available at   \texttt{http://nebel.rc.fas.harvard.edu/epsilon}.\end{abstract}


\maketitle

\section{Introduction}

Dark matter (DM) annihilation or other new physics could inject electromagnetically interacting particles into the universe during its early history, with potentially wide-ranging observable consequences. In particular, injection of ionizing particles during the cosmic dark ages will increase the residual ionization fraction, broadening the last scattering surface and modifying the anisotropies of the cosmic microwave background (CMB) \cite{Adams:1998nr, Chen:2003gz, Padmanabhan:2005es}. Exquisitely sensitive measurements of the CMB by a range of experiments (e.g. WMAP \cite{2013ApJS..208...19H}, SPT \cite{2013ApJ...779...86S,2014ApJ...782...74H} and ACT \cite{Sievers:2013ica}), and most recently by the \emph{Planck} satellite \cite{Planck:2015xua}, can thus place robust and model-independent constraints on such energy injections (e.g. \cite{Zhang:2007zzh, Galli:2009zc, Slatyer:2009yq, Kanzaki:2009hf, Hisano:2011dc, Hutsi:2011vx,2011PhRvD..84b7302G, Finkbeiner:2011dx, 2013PhRvD..87l3513S, Galli:2013dna, Lopez-Honorez:2013cua, Madhavacheril:2013cna}).

For the specific case of DM annihilation, precise constraints can be placed under the assumption that the power \emph{deposited} to the gas is directly proportional to that \emph{injected} at the same redshift, with some efficiency factor $f_\mathrm{eff}$. More generally, a model- and redshift-dependent effective efficiency factor $f(z)$ can be defined as the ratio of deposited power to injected power at a given redshift (within some arbitrary volume element), even if much of the deposited power originates from energy injections at earlier times. It has been demonstrated \cite{Finkbeiner:2011dx} that given a set of $f(z)$ functions characteristic of Weakly Interacting Massive Particle (WIMP) models for DM \cite{Slatyer:2009yq}, the impact on the CMB is identical at the sub-percent level, up to an overall normalization factor, suggesting that approximating $f(z)$ by a constant $f_\mathrm{eff}$ is reasonable.

In order to convert from deposited power into perturbations to the ionization history, the \emph{Planck} Collaboration \cite{Planck:2015xua} assumed a simple prescription for the fraction of deposited power proceeding to ionization. This prescription was set by the corresponding fraction for 3 keV electrons, as determined by detailed Monte Carlo studies of the interactions of such electrons by the gas \cite{Valdes:2008cr, Valdes:2009cq, 2010MNRAS.404.1869F,  MNR:MNR20624, Galli:2013dna}; we refer to this as the ``3 keV'' prescription. Earlier works often took an even simpler approach, the so-called ``SSCK'' prescription (based on work by Shull and van Steenberg \cite{1985ApJ...298..268S}, and Chen and Kamionkowski \cite{Chen:2003gz}), in which a fraction $(1 - x_e)/3$ of deposited power proceeds into ionization; here $x_e$ is the ionization fraction (as relevant to the gas species in question; if helium is neglected, this is just the hydrogen ionization fraction). In practice, these prescriptions do break down; however, since the ionizing energy produced by a given energy injection is set by the \emph{product} of the deposition-efficiency curve $f(z)$ and the fraction of deposited power proceeding to ionization, any errors in the prescription for the latter may be absorbed as corrections to $f(z)$. The \emph{Planck} Collaboration took this approach, employing corrected $f(z)$ curves, to set limits for several specific DM models: in this work we employ the same constraints, but demonstrate how to generalize them to a much broader class of DM models (all those where the velocity-weighted annihilation cross section $\langle \sigma v \rangle$ can be treated as constant during the cosmic dark ages).

In an accompanying article \cite{inprep}, referred to as Paper II, we have derived new results for the mapping from energy injection into power contributing to ionization, for $e^+e^-$ pairs and photons injected at arbitrary energy (in the keV to multi-TeV range) and arbitrary redshift (during the cosmic dark ages). Once an energy injection history is specified, the power proceeding into ionization at any redshift can be determined. In particular, in Paper II we have provided corrected $f(z)$ curves for DM annihilation (to photons and $e^+ e^-$ pairs) that compensate for the errors in the two most popular ionization prescriptions in the literature, ``SSCK'' and ``3 keV''. The only assumption on the annihilation model, in deriving these curves, is that the redshift dependence of the injected energy is $dE/dV dt \propto (1+z)^6$ (i.e. the power injected per unit volume is proportional to the square of the density), and that the spectrum of annihilation products is the same at all redshifts. To obtain the $f(z)$ curve for any given model, one need then only integrate over the spectrum of photons and $e^+ e^-$ pairs produced by the annihilation.

In this work we demonstrate that, in accordance with previous studies exploring a smaller range of DM models \cite{Finkbeiner:2011dx, Hutsi:2011vx,Madhavacheril:2013cna}, the impact of any conventional DM annihilation model on the CMB can be captured by a single normalization factor. The shape of the perturbation to the anisotropy spectrum is independent of the spectrum of annihilation products, and can be accurately obtained by assuming a constant redshift-independent $f(z)$. We derive and present the model-independent weighting function that determines the normalization factor, given a (corrected) deposition-efficiency curve $f(z)$. Having obtained this weighting function, we can convert the corrected $f(z)$ curves presented in Paper II directly into $f_\mathrm{eff}(E)$, for photons and $e^+ e^-$ pairs injected by DM annihilation at some arbitrary energy $E$. Computing $f_\mathrm{eff}$ for any conventional model of annihilating DM (i.e. one without a novel redshift-dependence) is then simply a matter of integrating over the spectrum of annihilation products. As an example, we perform this calculation for the  DM masses and Standard Model final states contained in \texttt{PPPC4DMID} \cite{Cirelli:2010xx}, and determine the resulting constraints from \emph{Planck} on the annihilation cross sections for these models.

We first review, in Sec. \ref{sec:review}, the interpretation of the corrected $f(z)$ curves that we employ. In Sec. \ref{sec:cmbbounds} we perform a principal component analysis (PCA) to derive the weighting function, construct $f_\mathrm{eff}$ as a function of injection energy and species, and demonstrate that the $f_\mathrm{eff}$ results are insensitive to various choices in the analysis. We demonstrate the application of these results to a range of DM models in Sec. \ref{sec:pppc}, and then present our conclusions in Sec. \ref{sec:conclusions}. The appendices provide information on supplemental data files, which we make available at \texttt{http://nebel.rc.fas.harvard.edu/epsilon}, and review some aspects of PCA and the derivation of the weighting function.

\section{The corrected $f(z)$ curves}
\label{sec:review}

We use the $f(z)$ curves from Paper II \cite{inprep}, and refer the reader to that work for a detailed description of their derivation. However, it is worth recapping the definition and properties of the three types of $f(z)$ curves we employ, which we will denote $f^\mathrm{SSCK}(z)$, $f^\mathrm{3keV}(z)$ and $f^\mathrm{sim}(z)$. In the notation of Paper II, these curves correspond to (respectively) $f^\mathrm{ion,SSCK}(z)$, $f^\mathrm{ion,3keV}(z)$ and $f^\mathrm{sim}(z)$. We will refer to these curves generally as $f(z)$ curves. 

The ``ion'' superscript indicates that the deposition-efficiency curves have been corrected to ensure that when used with the stated prescription (``SSCK'' or ``3 keV''), one recovers the correct total power into ionization. This is the appropriate condition for constraints arising from perturbations to the ionization history, which in turn modify the anisotropies of the CMB. It has been shown that in the case of annihilating DM, other deposition channels can essentially be ignored: for example, the limits from high-energy free-streaming photons (discussed briefly in e.g. \cite{Padmanabhan:2005es}), distortions to the CMB spectrum \cite{Zavala:2009mi, Hannestad:2010zt, 2012MNRAS.419.1294C, Chluba:2013wsa} and modifications to the gas temperature (e.g. \cite{Cirelli:2009bb, Lopez-Honorez:2013cua}) are all much weaker (for $s$-wave-annihilating DM), and ignoring production of Lyman-$\alpha$ photons altogether only changes the constraints by $\sim 5$\% \cite{Hutsi:2011vx, Galli:2013dna}. Accordingly, in this paper we will only ever use $f(z)$ curves corrected to obtain an accurate description of the power proceeding into ionization (in preference to the other deposition channels), and the ``ion'' superscript is unnecessary.

The $f^\mathrm{sim}(z)$ curve is obtained from a simplified estimate for the power deposited into each channel. The overall deposited power is corrected by subtracting previously-unaccounted power lost into photons below 10.2 eV, which are no longer sufficiently energetic to interact efficiently with the gas (see Paper II \cite{inprep} for details), and then the remainder is multiplied by the ``SSCK'' or ``3 keV'' prescription to obtain the power into ionization. This approach is denoted ``approx'' in \cite{Galli:2013dna}, in contrast to the ``best'' treatment that yields the deposition fractions which the $f^\mathrm{SSCK}(z)$ and $f^\mathrm{3keV}(z)$ functions are designed to reproduce; it relies on the assumption that the ``SSCK'' or ``3 keV'' prescriptions are accurate once the energy losses of higher-energy electrons (above 3 keV) to very low-energy photons are taken into account. However, as shown for a range of specific models in \cite{Galli:2013dna}, this simplified approach gives results comparable to the more detailed calculation.

\begin{figure*}
\includegraphics[width=0.305\textwidth]{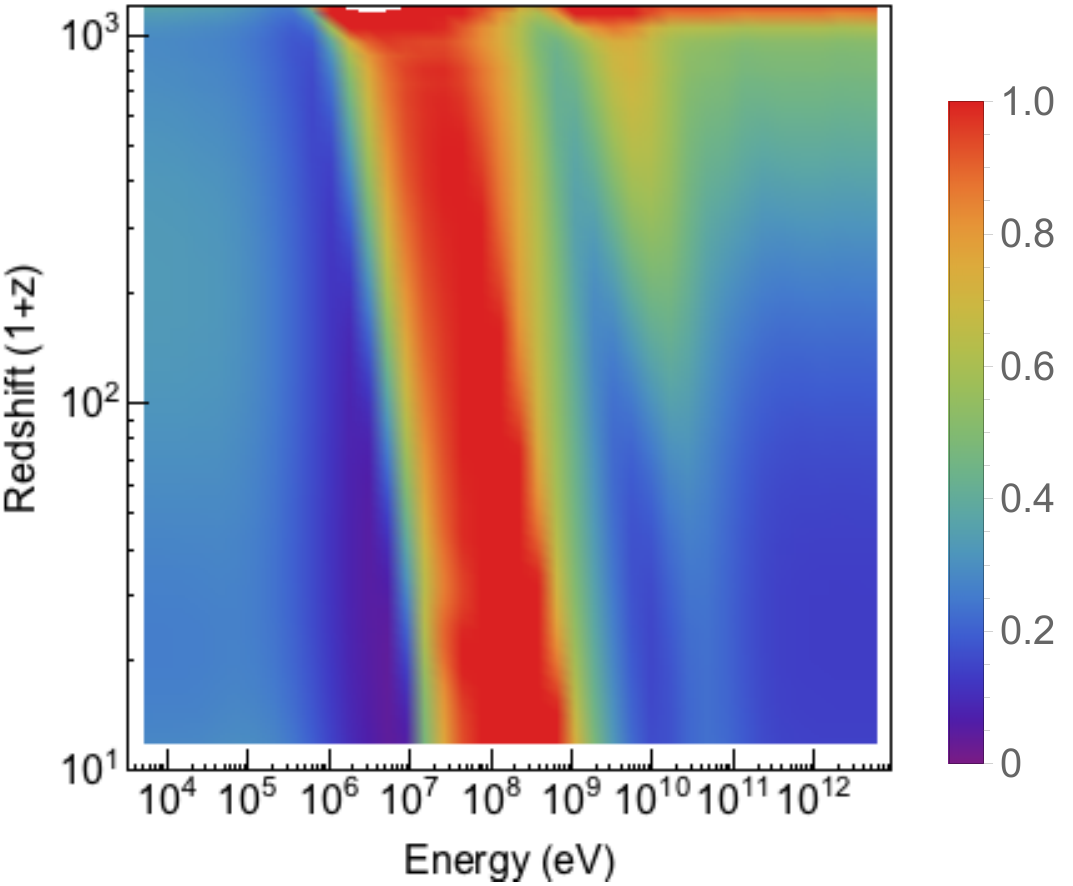}
\includegraphics[width=0.305\textwidth]{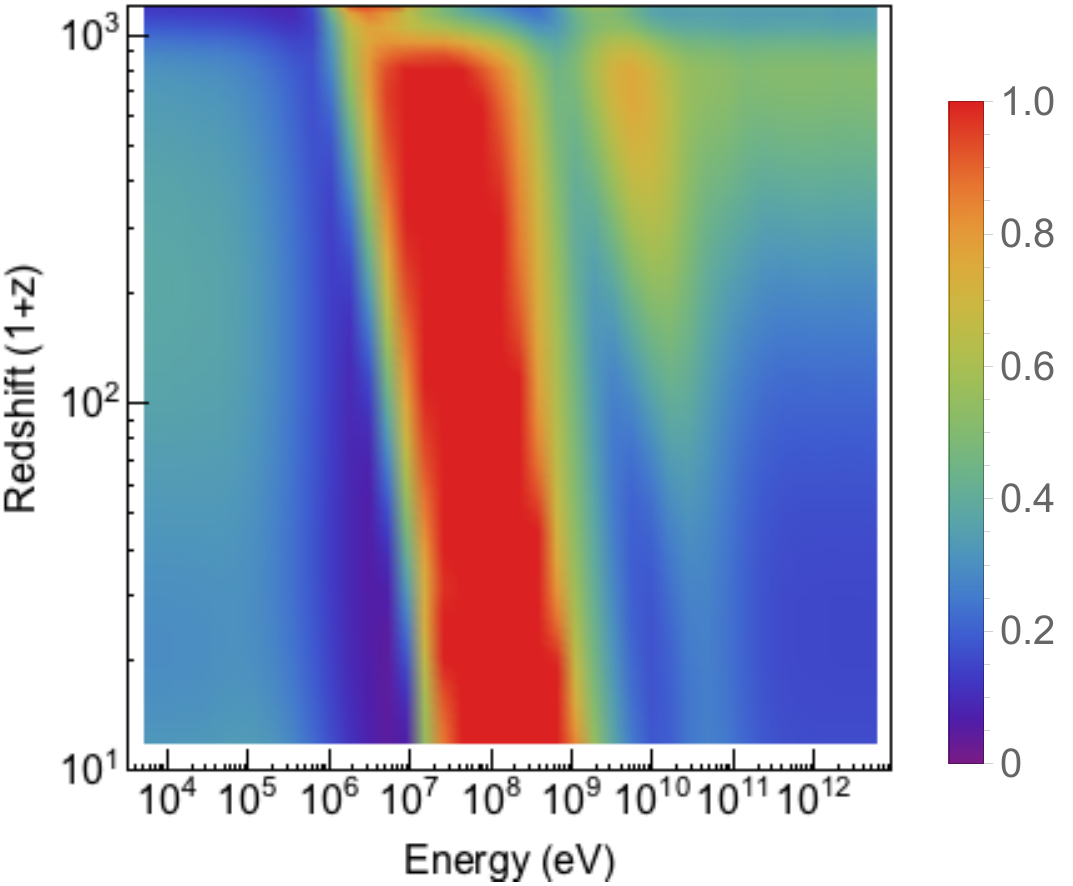}
\includegraphics[width=0.31\textwidth]{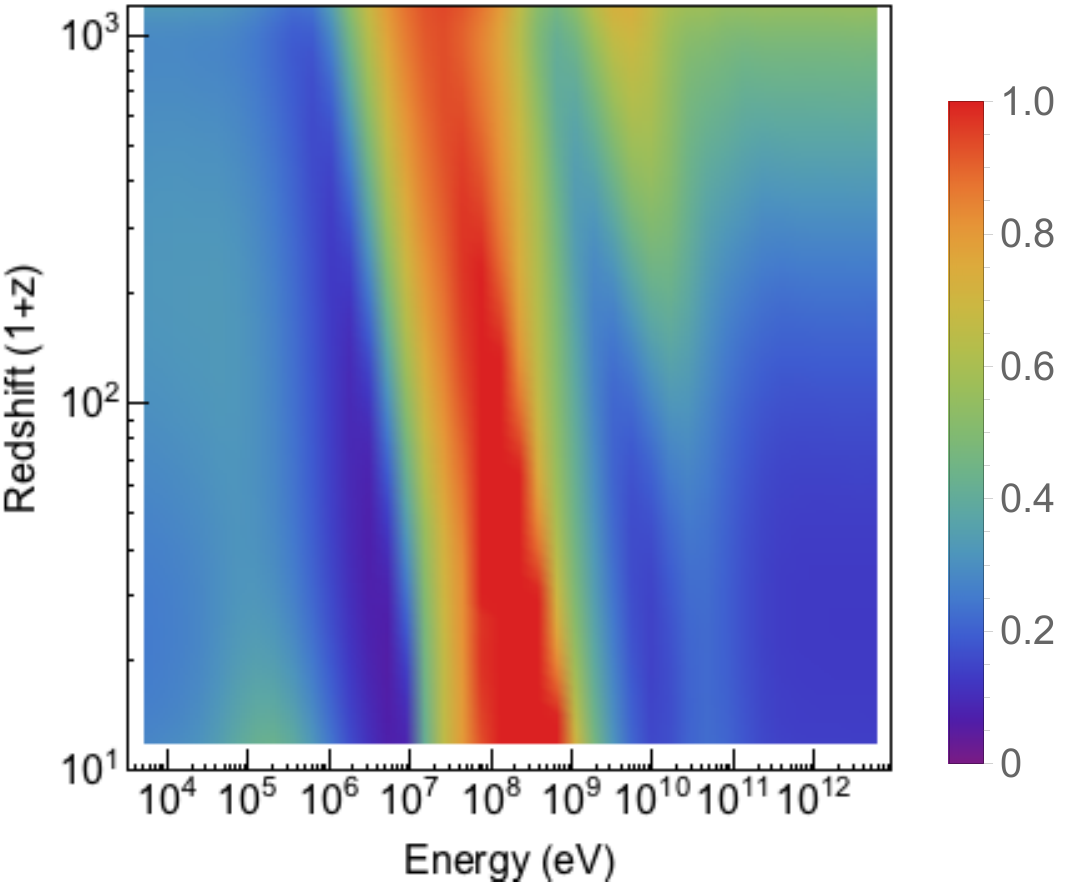} \\
\includegraphics[width=0.305\textwidth]{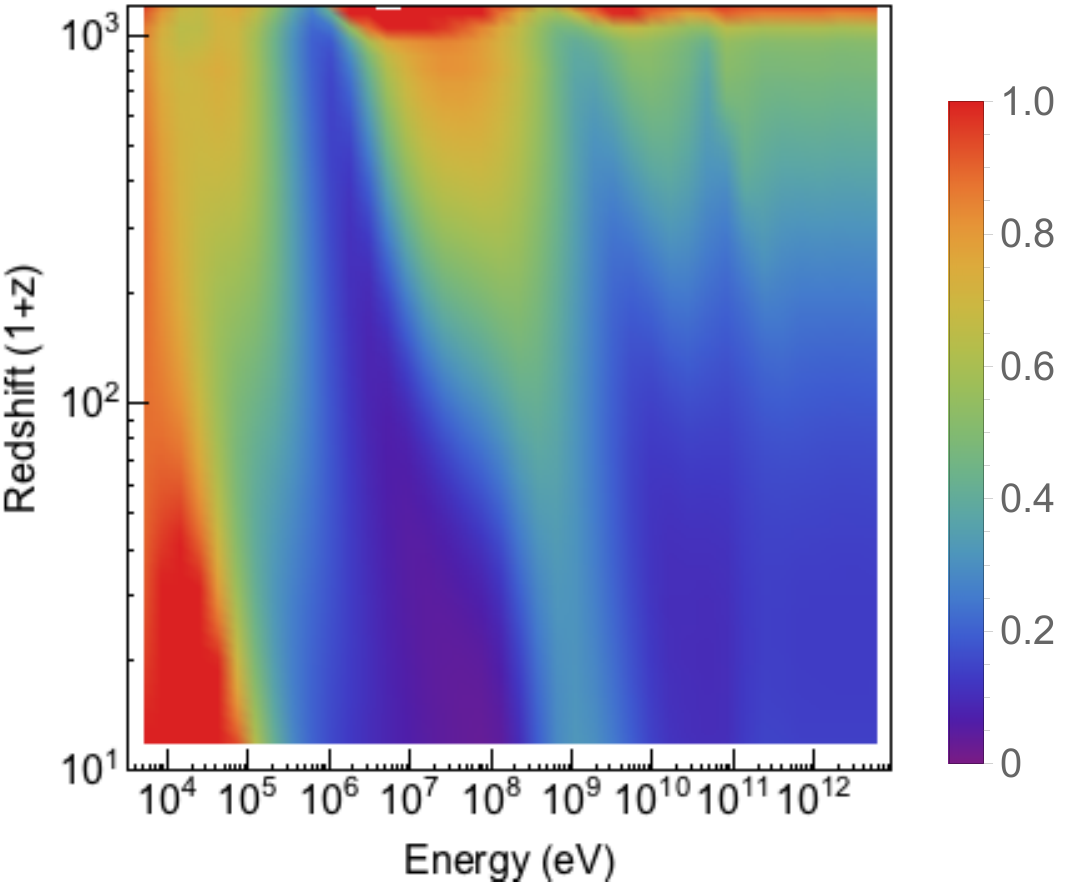}
\includegraphics[width=0.305\textwidth]{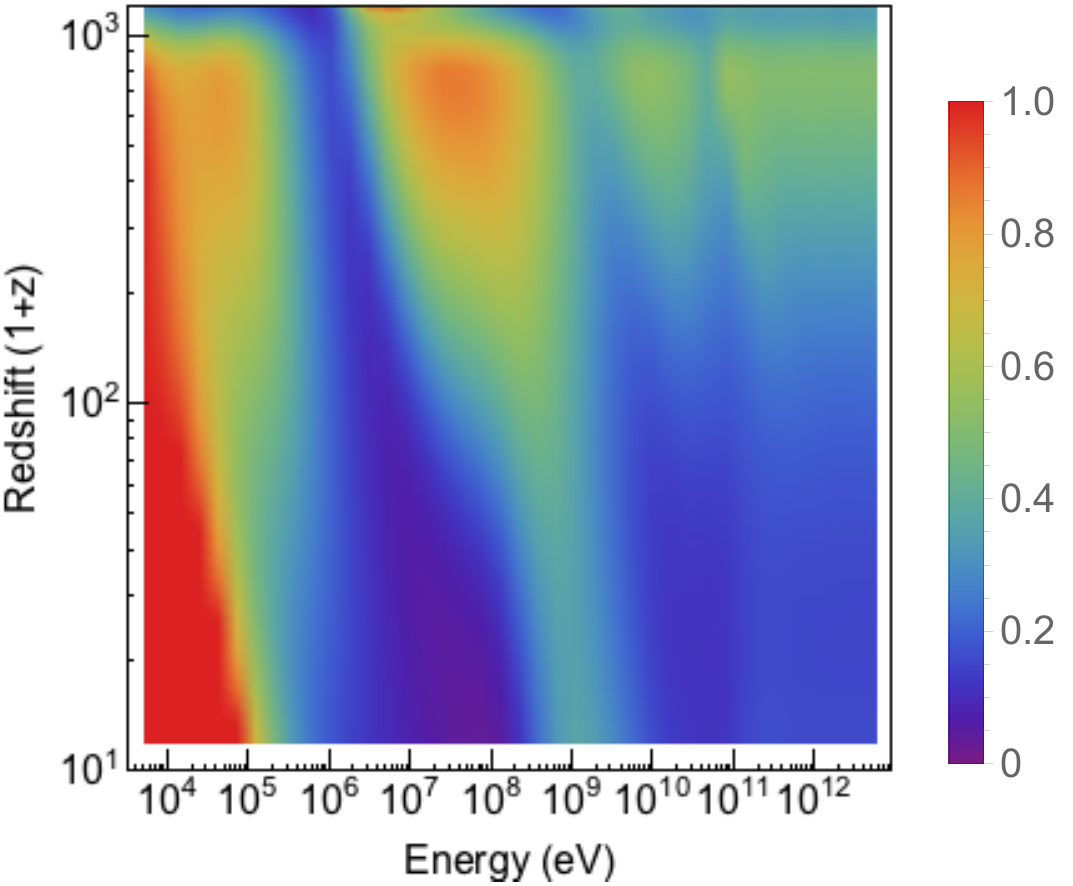}
\includegraphics[width=0.31\textwidth]{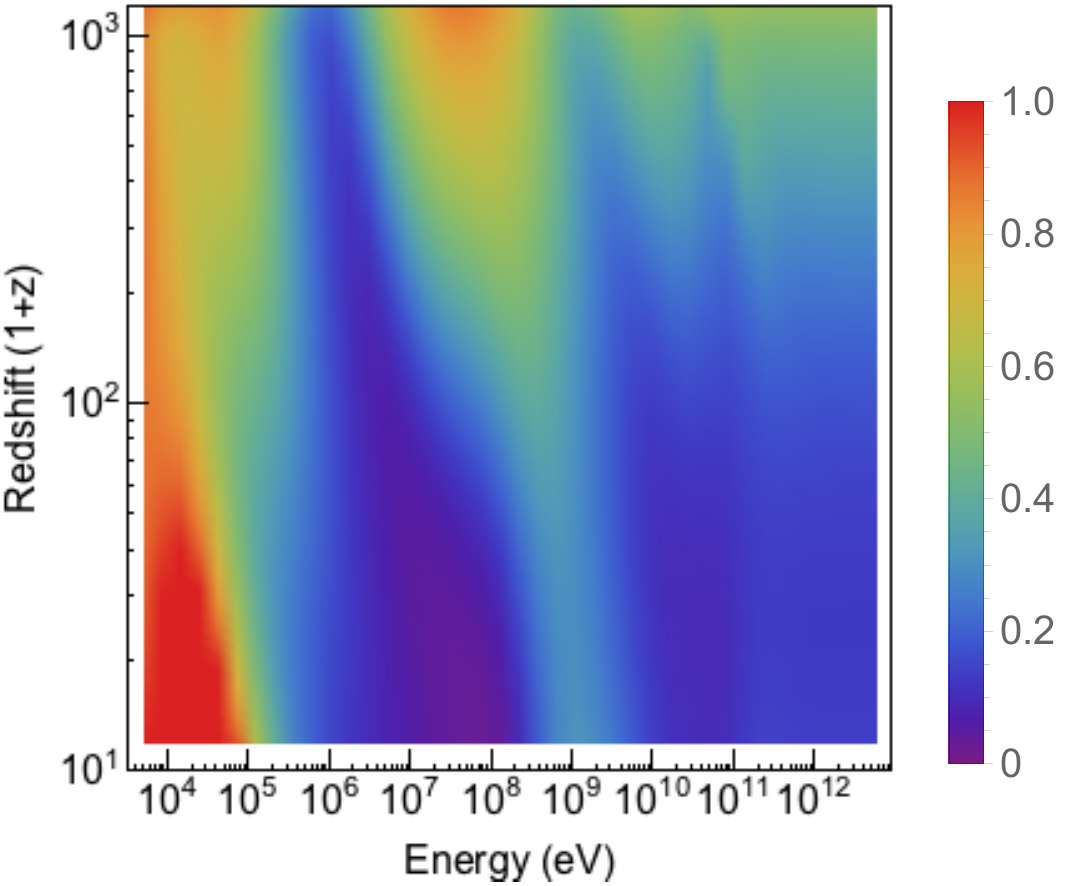}
\caption{\label{fig:fDM}
Corrected $f(z)$ functions for particles injected by DM annihilation, as a function of injection energy and redshift of absorption. In the left panel we use the ``3 keV'' baseline ionization fractions (so these $f^\mathrm{3keV}(z)$ curves should be used with analyses that employed the same prescription); in the center panel we use the ``SSCK'' baseline. In the right panel we plot the simplified channel-independent $f^\mathrm{sim}(z)$ curve. The upper row describes $e^+ e^-$ pairs (the $x$-axis ``energy'' label here indicates the kinetic energy of a single member of the pair at injection), the lower row describes photons.
}
\end{figure*}

To summarize, the $f^\mathrm{SSCK}(z)$, $f^\mathrm{3keV}(z)$ are our best estimates for the appropriate corrected $f(z)$ curves to use with studies that assumed the ``SSCK'' or ``3 keV'' prescriptions, respectively; the $f(z)$ curves have been corrected to take into account errors in those prescriptions. The $f^\mathrm{sim}(z)$ curve is the result of a simpler calculation and can be used as a cross-check. We plot all three sets of curves in Fig. \ref{fig:fDM}, as taken from Paper II \cite{inprep}. Note that all these curves only account for the energy injected by annihilation of the smooth component of DM with the cosmological average density; they do not take into account structure formation at low redshifts. However, it has been shown that the contribution from structure formation to the CMB constraints is generally very subdominant for $s$-wave annihilation \cite{Lopez-Honorez:2013cua}.

\section{Converting corrected $f(z)$ curves to CMB bounds}
\label{sec:cmbbounds}

The most recent and strongest constraints on DM annihilation from CMB anisotropies come from an analysis of \emph{Planck} data \cite{Planck:2015xua}. These constraints were computed using the ``3 keV'' baseline prescription, and were expressed in terms of a bound on $f_\mathrm{eff} \langle \sigma v \rangle / m_\mathrm{DM}$. For the models tested by those authors, $f_\mathrm{eff}$ was determined by evaluating the appropriate corrected $f(z)$ curve for the model at $z=600$, as originally suggested and validated in \cite{2011PhRvD..84b7302G}. Analogous $f_\mathrm{eff}$ values for arbitrary photon and $e^+ e^-$ energies can be read off directly from Fig. \ref{fig:fDM}, using the ``3 keV'' baseline prescription or the simplified $f^\mathrm{sim}(z)$ curves (as discussed above, the latter approach is simplified and likely to be less accurate, but it gives a rough estimate of the possible systematic errors arising from details of the first method).

 A more careful derivation of $f_\mathrm{eff}$ -- that is, the \emph{constant} value of $f(z)$ which would have the same impact on the CMB as the true corrected $f(z)$ curve -- can be achieved by PCA, following the approach of \cite{Finkbeiner:2011dx}.\footnote{For convenience, in this work we use the marginalized Fisher matrices derived and made publicly available by \cite{Finkbeiner:2011dx}; while those results employed WMAP measurements of the cosmological parameters to determine the fiducial cosmological model, we expect  the impact of updating to \emph{Planck} cosmological parameters to be extremely small.} We review the essentials of the Fisher matrix and PCA in Appendix \ref{app:pca}. Performing PCA on a range of $f(z)$ curves corresponding to assorted DM models, \cite{Finkbeiner:2011dx, Madhavacheril:2013cna} found that the vast majority of the variance -- exceeding $99.9\%$ -- was contained in the first principal component, peaking around $z\sim 600$. Related studies \cite{2011PhRvD..84b7302G,Hutsi:2011vx,2012JCAP...12..008G} independently determined, using different methods, that the signal was largely controlled by a single parameter, set by the behavior of $f(z)$ at $z \sim 600$. 
 
 As described in \cite{Madhavacheril:2013cna}, the ``dot product'' of a given $f(z)$ curve with the first principal component thus completely determines its detectability. This dot product can be re-expressed as an integral over $d\ln(1+z)$, with the integrand being the product of the $f(z)$ curve with a weighting function $W(z)$. The weighting function can be determined from the first principal component, as described in \cite{Finkbeiner:2011dx, Madhavacheril:2013cna}; we provide a detailed derivation in Appendix \ref{app:weightfn}. (An alternate PCA, based on the ionization history rather than the energy deposition history, was performed in \cite{2012ApJ...752...88F}.)
 
In principle, the weighting function $W(z)$ depends on the details of the experiment (specifically, its sensitivity to different multipoles $\ell$, in both temperature and polarization). It will also depend on the ionization prescription, as this affects the mapping between $f(z)$ and the CMB. In practice, however, the impact of these choices turns out to be very small.

We performed a new PCA, building our model space from the $f(z)$ curves for annihilation to $e^+e^-$ and photons, for 20 log-spaced energies between 1 keV and 10 TeV (for 40 models total). This is considerably more general than previous studies which focused on the GeV-TeV WIMP parameter space, but again we find that the first principal component overwhelmingly dominates the variance (accounting for more than $99.7\%$ in all cases). In other words, the space spanned by the possible imprints of conventional $s$-wave DM annihilation on the CMB is (within the confines of the approximate Fisher-matrix approach) essentially one-dimensional.

In order to justify computing the constraints on constant $f(z)$ and applying them to arbitrary annihilation models, we need to show that scenarios with constant $f(z)$ also produce a perturbation to the CMB anisotropies lying within this one-dimensional space. To test this question, we consider two energy injection histories, one with $f(z) =$ constant, the other with $f(z)$ given by the first principal component, with the same DM mass and annihilation cross section. Suppose the $f(z)$ curves are normalized such that they correspond to the same signal significance. As discussed in Appendix \ref{app:pca} (and using the notation defined there), if $\vec{v}_1$ and $\vec{v}_2$ are the discretized version of the $f(z)$ curves, this condition implies $\vec{v}_1^T F_z \vec{v}_1 = \vec{v}_2^T F_z \vec{v}_2$, where $F_z$ is the Fisher matrix  (as determined in \cite{Finkbeiner:2011dx}).

The significance of the difference between these energy injection histories, normalized to the overall significance of either history, can be estimated within the Fisher matrix approach as:
\begin{equation}\sigma_\Delta = \sqrt{ \frac{(\vec{v}_2 - \vec{v}_1)^T F_z (\vec{v}_2 - \vec{v}_1)}{\vec{v}_1^T F_z \vec{v}_1}} = \sqrt{2 \left(1 - \frac{\vec{v}_2^T F_z \vec{v}_1}{\vec{v}_1^T F_z \vec{v}_1} \right)}.  \label{eq:dotproduct}\end{equation}
Here we have used the normalization condition above, and also the symmetry of the Fisher matrix to write $\vec{v}_1^T F_z \vec{v}_2 =  \vec{v}_2^T F_z \vec{v}_1$. We computed the bracketed quantity on the RHS of Eq. \ref{eq:dotproduct}, taking $f_1(z)$ to be constant and $f_2(z)$ to be the first principal component. We found that in all cases the bracketed quantity was below $0.001$, indicating that to a good approximation, the first principal component also captures the effect of constant $f(z)$. Thus it is reasonable to derive constraints on the case of constant $f(z)$ and convert them to constraints on general DM annihilation histories using the weighting function.

We repeated the analysis for each independent combination of the following choices:
\begin{itemize}
\item A \emph{Planck}-like experiment vs a cosmic variance limited (CVL) experiment (as defined in \cite{Finkbeiner:2011dx}).
\item Including multipoles up to $\ell =2500$, vs $\ell=6000$.
\item Using $f^\mathrm{SSCK}(z)$, $f^\mathrm{3keV}(z)$ and $f^\mathrm{sim}(z)$ to generate the space of $f(z)$ curves.
\item Choosing ``SSCK'' vs ``3 keV'' as our baseline ionization prescription (to translate the $f(z)$ curve into its impact on the CMB) when using the $f^\mathrm{sim}(z)$ curve.
\end{itemize}
In all cases we used a Fisher analysis to estimate detectability in the CMB, as described in \cite{Finkbeiner:2011dx}. We normalized the weighting function as described in \cite{Madhavacheril:2013cna} and derived in detail in Appendix \ref{app:weightfn}, so that $\int W(z) d\ln(1+z) = 1$. Thus for any arbitrary $f(z)$, $\int f(z) W(z) d\ln(1+z)$ measures detectability relative to the case of $f(z)=1$.

\begin{figure*}
\includegraphics[width=0.32\textwidth]{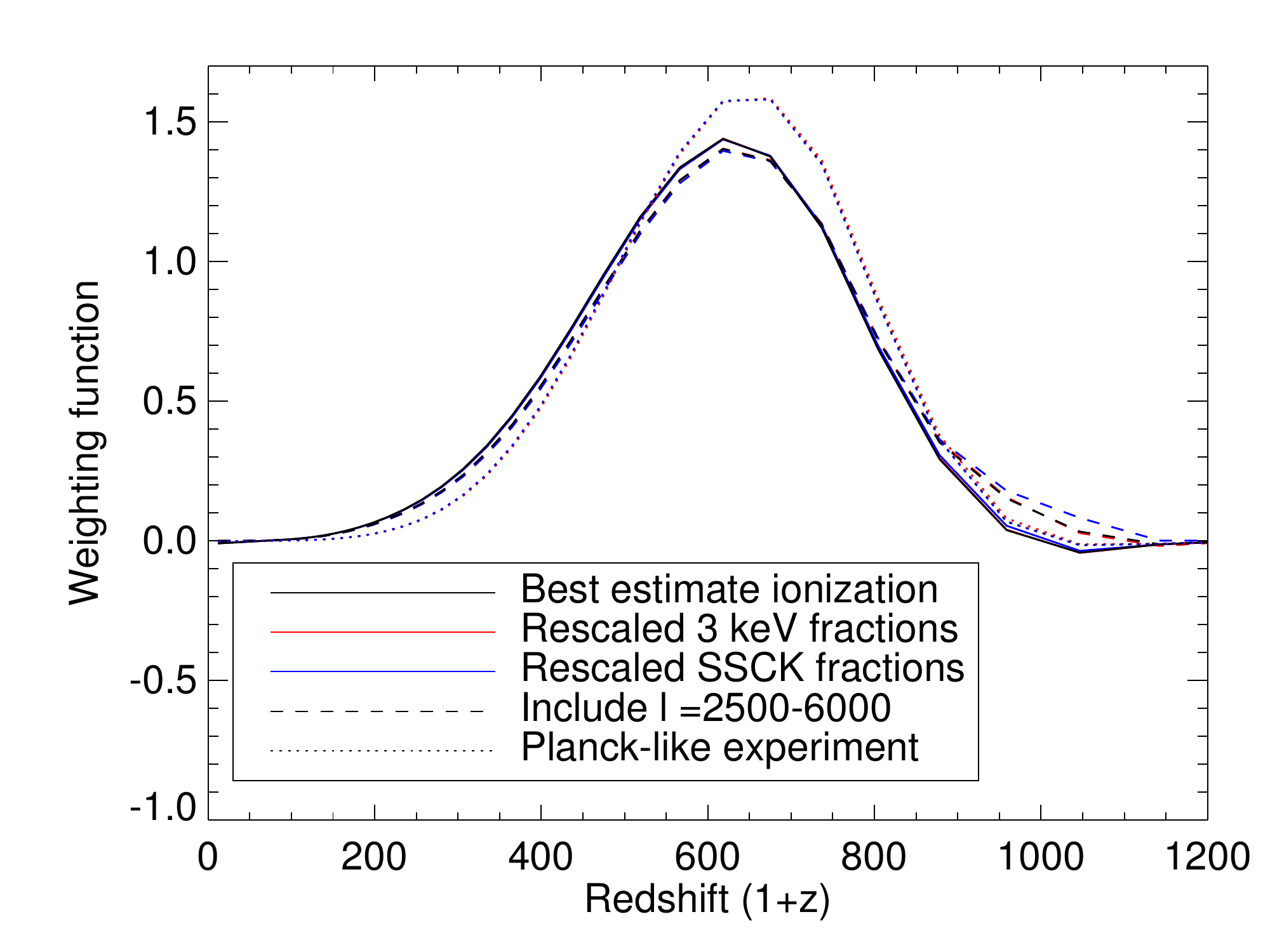}
\includegraphics[width=0.32\textwidth]{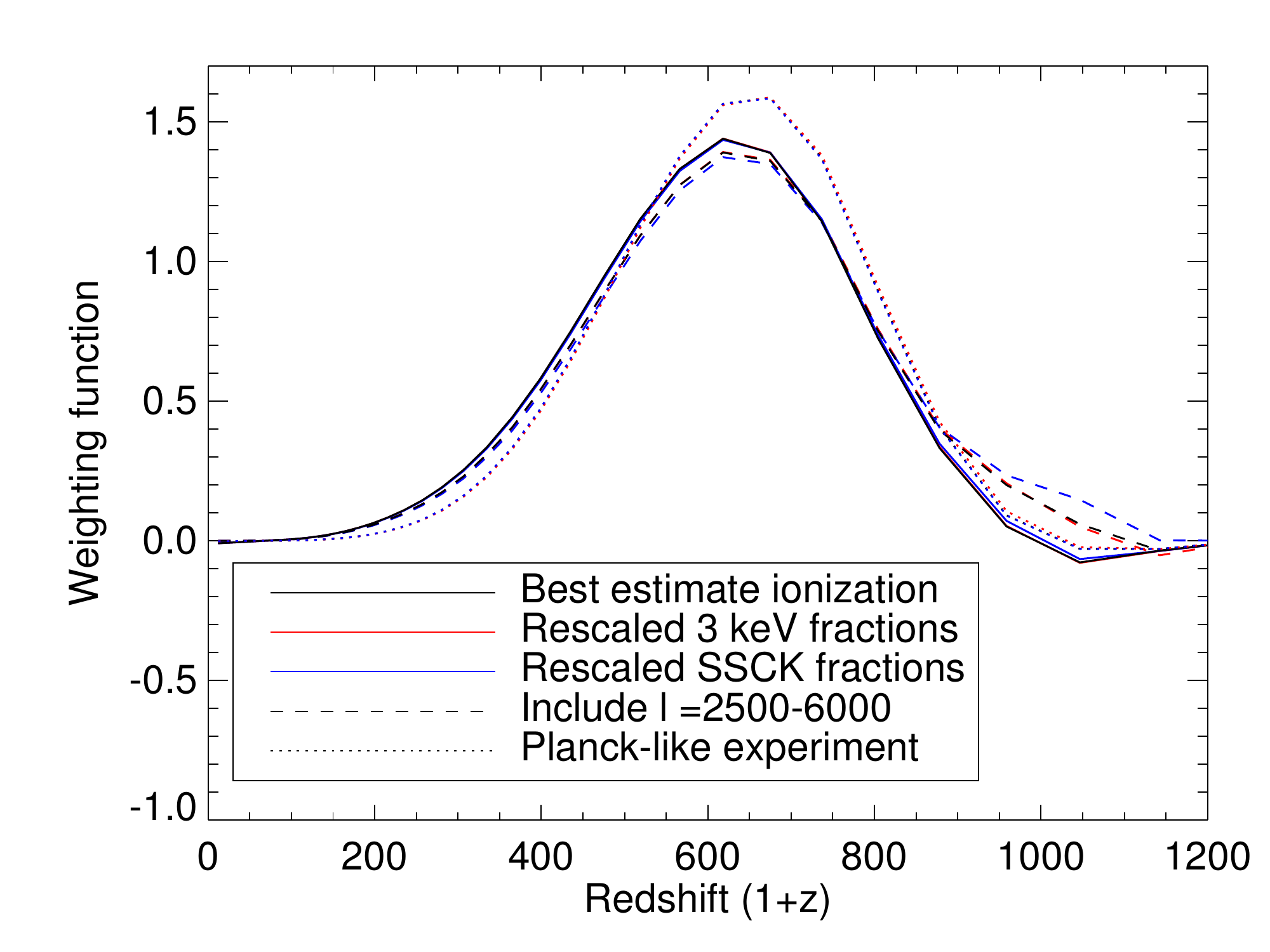}
\includegraphics[width=0.32\textwidth]{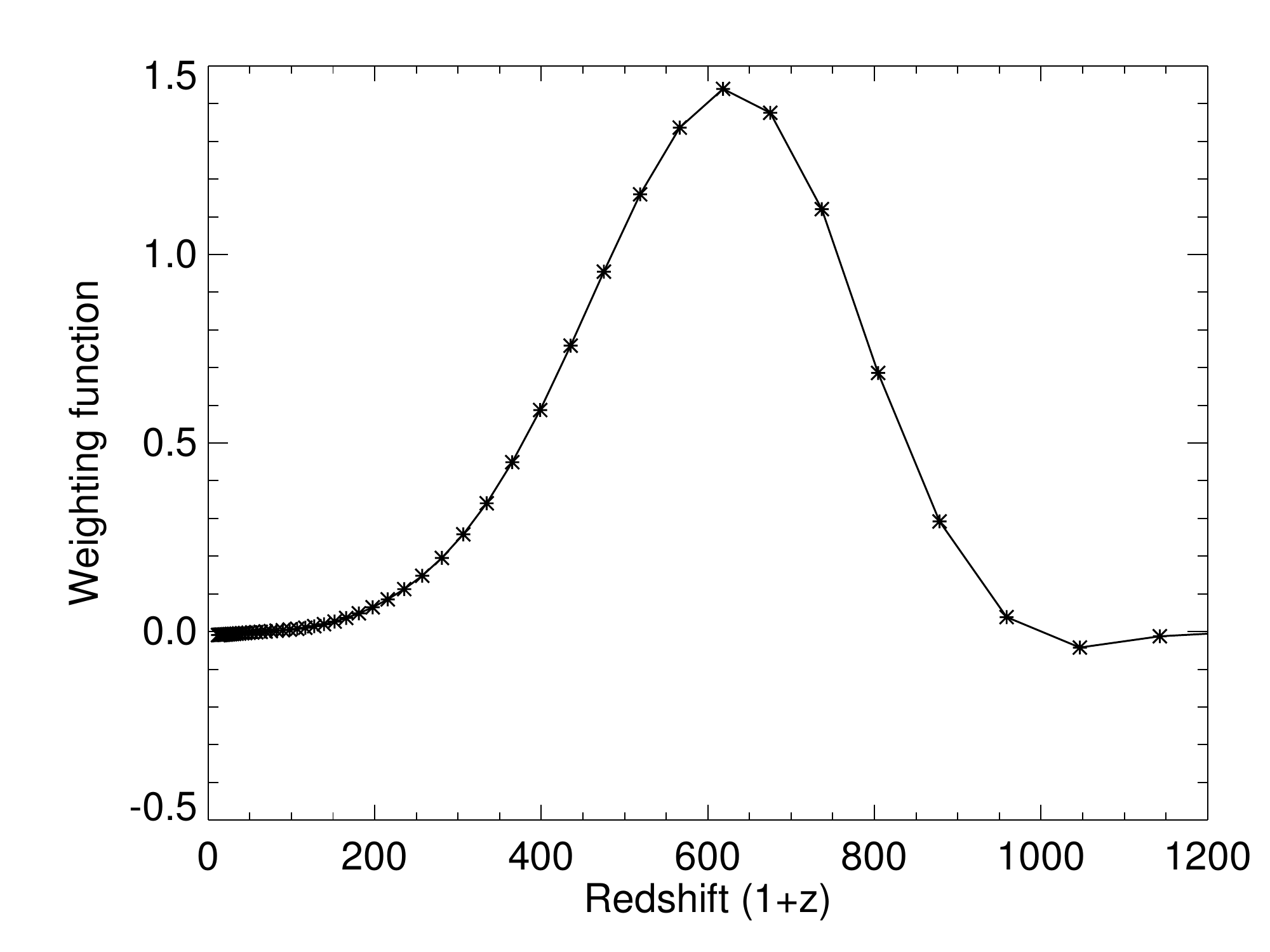}
\caption{\label{fig:weightingfn}
Weighting functions for the $f(z)$ curves, derived by PCA. Solid lines show the results for a CVL experiment covering $\ell=2-2500$, dashed lines show the results for a CVL experiment covering $\ell=2-6000$, dotted lines show the result for a \emph{Planck}-like experiment covering $\ell=2-2500$. \emph{Black} lines use the first procedure for estimating ionization history, \emph{red} lines use the second procedure with ``3 keV'' baseline prescription, \emph{blue} lines use the second procedure with ``SSCK'' baseline prescription. Left panel: suitable for ``3 keV'' baseline. Center panel: suitable for ``SSCK'' baseline. Right panel: ``universal'' weighting function recommended for general use.
}
\end{figure*}

\begin{figure*}
\includegraphics[width=0.45\textwidth]{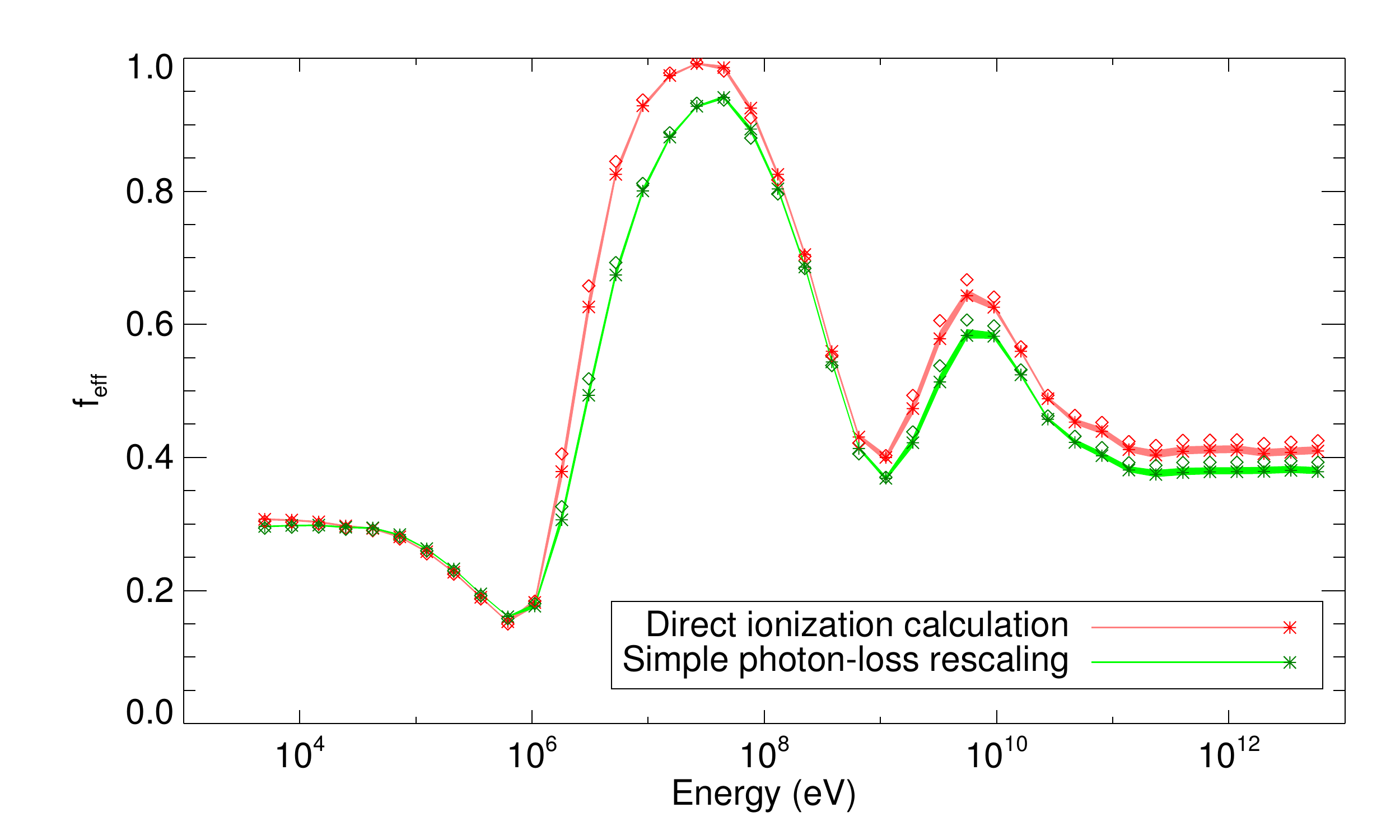}
\includegraphics[width=0.45\textwidth]{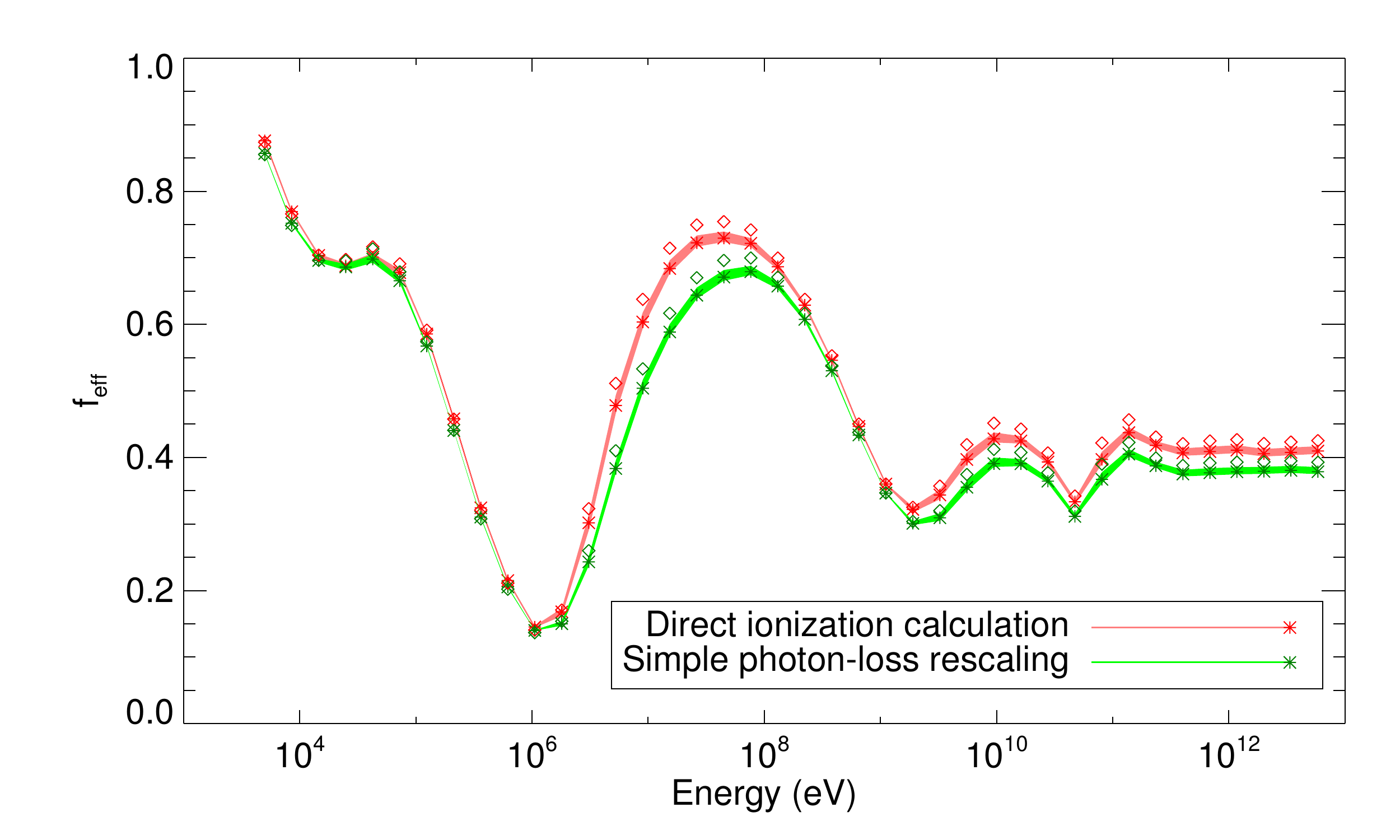} \\
\includegraphics[width=0.45\textwidth]{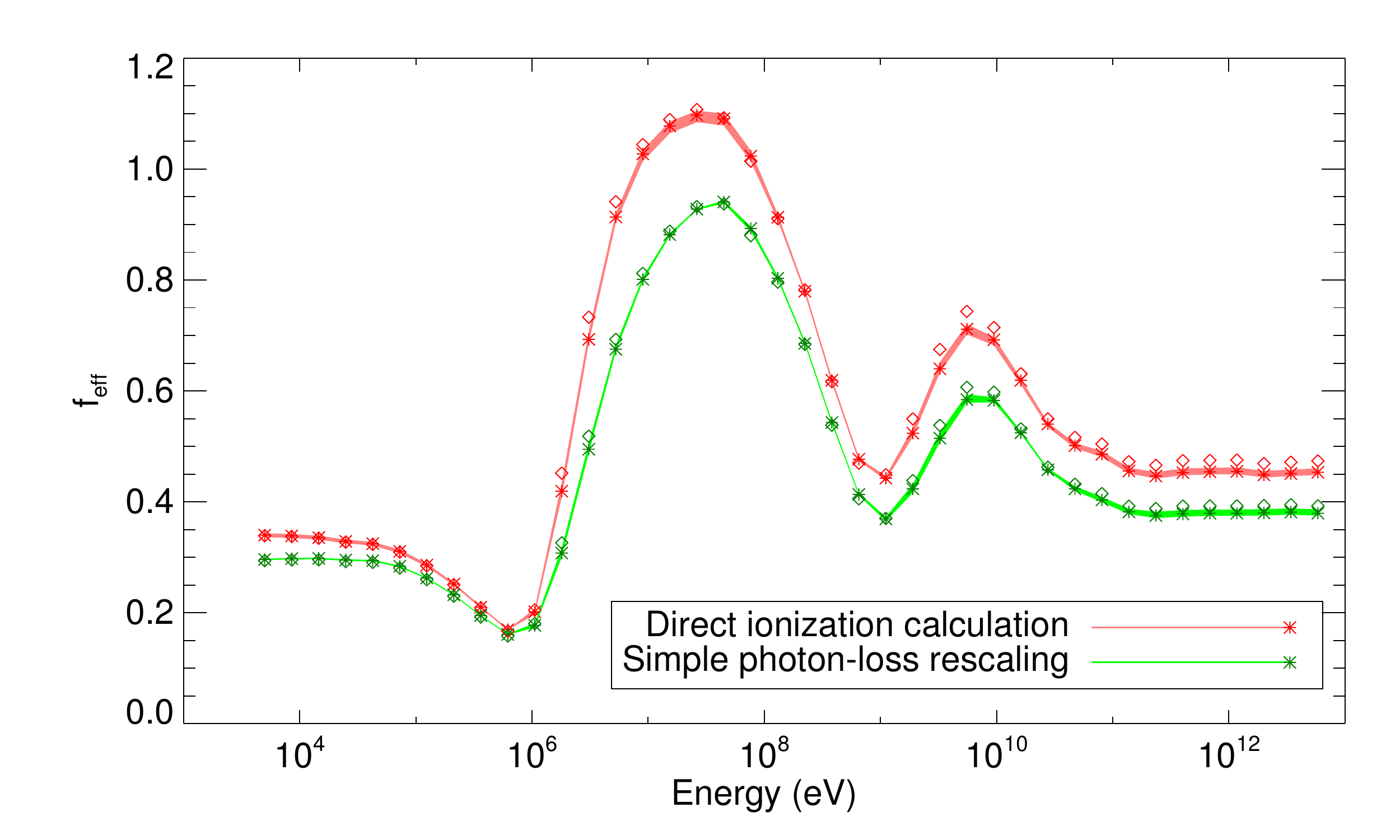}
\includegraphics[width=0.45\textwidth]{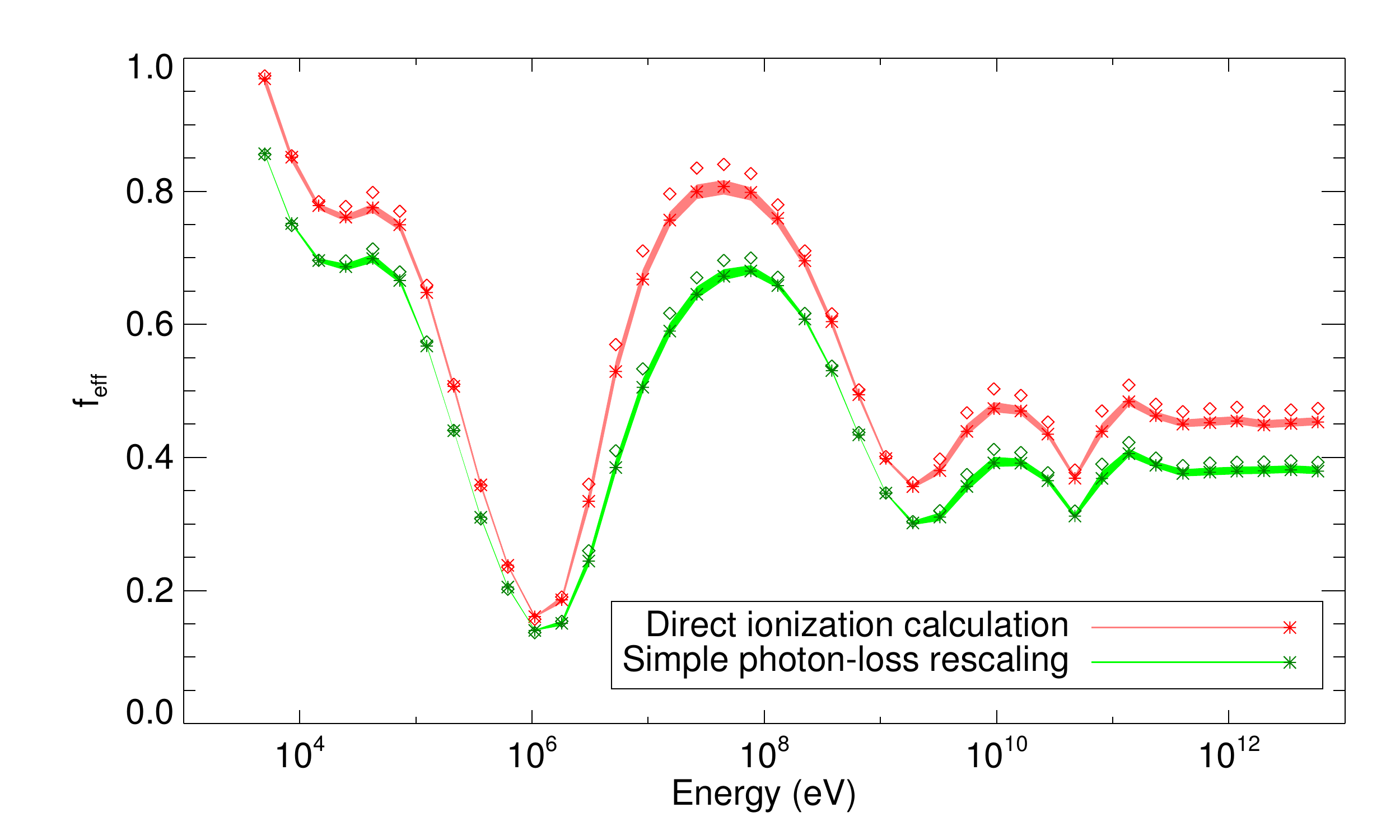}
\caption{\label{fig:feff}
$f_\mathrm{eff}$ coefficients as a function of energy for $e^+ e^-$ (left column) and photons (right column); appropriate for analyses with baseline prescription ``3 keV'' (top row) or ``SSCK'' (bottom row). The widths of the red and green bands indicate the impact of scanning over all the derived weighting functions (see Fig. \ref{fig:weightingfn}). Red stars are derived from the $f^\mathrm{SSCK}(z)$ or $f^\mathrm{3keV}(z)$ curves as appropriate, green stars from the $f^\mathrm{sim}(z)$ curves.  Diamonds indicate $f_\mathrm{eff}$ evaluated as $f(600)$, rather than using a weighting function, and are labeled ``simple photon-loss rescaling'' in the legend.
}
\end{figure*}

There are visible differences in the weighting functions thus derived, with the largest impacts coming from changing the sensitivity to different multipoles -- that is, by considering a \emph{Planck}-like experiment, vs a CVL experiment up to $\ell=2500$, vs a CVL experiment up to $\ell=6000$. However, the resulting variation in $f_\mathrm{eff}$ computed by the different $W(z)$ functions is always negligible (percent-level or less). We show the various $W(z)$ curves in Fig. \ref{fig:weightingfn}; in Fig. \ref{fig:feff} we display the impact on the $f_\mathrm{eff}$ fractions of scanning over all derived $W(z)$ curves.

We recall that if the $f^\mathrm{SSCK}(z)$ and $f^\mathrm{3keV}(z)$ curves are each used with their specified ionization prescription, they correspond to identical changes to the ionization history. Consequently, their impacts on the CMB must be the same (if contributions from non-ionization channels are ignored). However, since the $f(z)$ curves are different, one might expect that the weighting function $W(z)$ will be different -- and so must be chosen to match the ionization prescription under which it was derived -- in order to compensate.  We discuss this point in detail in Appendix \ref{app:ionpca}, but it turns out that in practice, the similarity of the different ionization prescriptions and the unit normalization of $W(z)$ means the effect is tiny. Instead, the difference in baseline ionization prescriptions is captured in the $f_\mathrm{eff}$ parameter, and the derived constraints on that parameter; for example, a higher baseline prescription for ionization would correspond to lower $f(z)$ for all models, and hence lower $f_\mathrm{eff}$, but the constraints on a given $f_\mathrm{eff}$ would be stronger (since the higher ionization prescription corresponds to a larger effect on the CMB). 

We have explicitly tested the effect of using the ``wrong'' weighting function -- i.e. using $W(z)$ derived using the $f^\mathrm{3keV}(z)$ curves, in order to estimate $f_\mathrm{eff}$ for $f^\mathrm{SSCK}(z)$ curves. The effect is included in the bands in Fig. \ref{fig:feff}, and is negligible.

Consequently, for the purposes of computing $f_\mathrm{eff}$, it is entirely adequate to use a single weighting function; we choose the weighting function derived for a CVL experiment including $\ell$ up to 2500, built from the $f^\mathrm{3keV}(z)$ curves. We note that this choice is largely arbitrary, as the resulting value of $f_\mathrm{eff}$ is very stable. This weighting function is likely to be broadly applicable beyond the case of DM annihilation, although it may not capture all the variance for energy injection histories that differ markedly from that of conventional DM annihilation. We leave a more detailed PCA, based on a larger range of possible injection histories, for future work.

One can integrate any electron and photon spectra produced by DM annihilation over the $f_\mathrm{eff}$ curves presented in Fig. \ref{fig:feff}, in order to determine $f_\mathrm{eff}$ for an arbitrary model. We show results separately for the ``3 keV'' and ``SSCK'' baseline prescriptions; however, when making comparisons with the \emph{Planck} constraints \cite{Planck:2015xua}, the ``3 keV'' baseline prescription should always be used. We also show the effect of taking our second simplified procedure for computing $f_\mathrm{corr}(z)$, and the effect of evaluating $f_\mathrm{corr}(600)$ rather than using the weighting function; both effects are rather small, at the $< 10\%$ level.

The $f_\mathrm{eff}$ curves (Fig. \ref{fig:feff}) exhibit significant structure as a function of energy. At high energies, both $e^+ e^-$ pairs and photons asymptote to $f_\mathrm{eff} \approx 0.4$; this convergent behavior is expected, since at high energies electrons and photons both participate in a pair production / ICS cascade. Both species demonstrate some minor structure in the GeV-TeV range but generally maintain $f_\mathrm{eff} \sim 0.4$, with a peak of $f_\mathrm{eff} \sim 0.7$ for electrons at injection energies around 10 GeV. Both species then exhibit a large bump in $f_\mathrm{eff}$, peaking at energies around 10-100 MeV, and a trough at $\mathcal{O}$(MeV) energies. This trough occurs because mildly relativistic electrons suffer large losses into continuum photons with energies below 10.2 eV (see Paper II \cite{inprep} for an in-depth discussion). Accordingly, MeV-scale electrons have small $f_\mathrm{eff}$, and the dominant energy loss process for MeV-scale photons is Compton scattering, which in turn produces more mildly-relativistic electrons. The peak at slightly higher energy occurs because 10-100 MeV electrons lose most of their energy by inverse Compton scattering into $\mathcal{O}(100-10^4)$ eV photons, which are efficient ionizers. 

At low energies, below $\sim 1-100$ keV (depending on the redshift), photons become efficient photoionizers, causing a large $f_\mathrm{eff}$. In contrast, the $e^+ e^-$ $f_\mathrm{eff}$ at low kinetic energies is rather low, of order 0.3 -- but as noted in \cite{2013PhRvD..87l3513S}, this is because in the context of DM annihilation, almost all the injected power is bound up in the mass energy of the pair in this case. Consequently, the absorption of the kinetic energy is almost irrelevant: what matters is the absorption of the photons produced when the positron annihilates. From inspection of $f_\mathrm{eff}$ for $\sim 0.5$ MeV photons, one would expect $f_\mathrm{eff}$ of 0.3 for pairs produced nearly at rest, consistent with the results of the calculation.

The primary systematic uncertainties in this calculation arise from the choice of $f(z)$ curves ($f^\mathrm{SSCK, 3keV}(z)$ or $f^\mathrm{sim}(z)$). Both of these curves involve an approximation; on one hand matching the power into ionization exactly (up to the limitations of our approximate treatment of the low-energy photons) while not attempting to match the heating and Lyman-$\alpha$ channels, on the other approximating \emph{all} channels by the cooling of 3 keV electrons after accounting for energy lost to continuum photons. Some approximation of this form is inevitable, given that the true relative losses into the different channels are \emph{not} independent of the injected particle energy as tacitly assumed by both the ``SSCK'' and ``3 keV'' prescriptions. As discussed more extensively in \cite{Galli:2013dna}, we estimate the potential systematic error due to these approximations and our simplified treatment of the low-energy photons to be at the $\sim 10\%$ level. For some channels, the constraints we present may also be slightly too conservative, i.e. weaker than the true limits, at a similar $\sim 10\%$ level, due to our neglect of energy deposition by protons and antiprotons \cite{Weniger:2013hja}.

\section{Constraints for simple annihilation channels}
\label{sec:pppc}

\begin{figure*}
\includegraphics[width=0.5\textwidth]{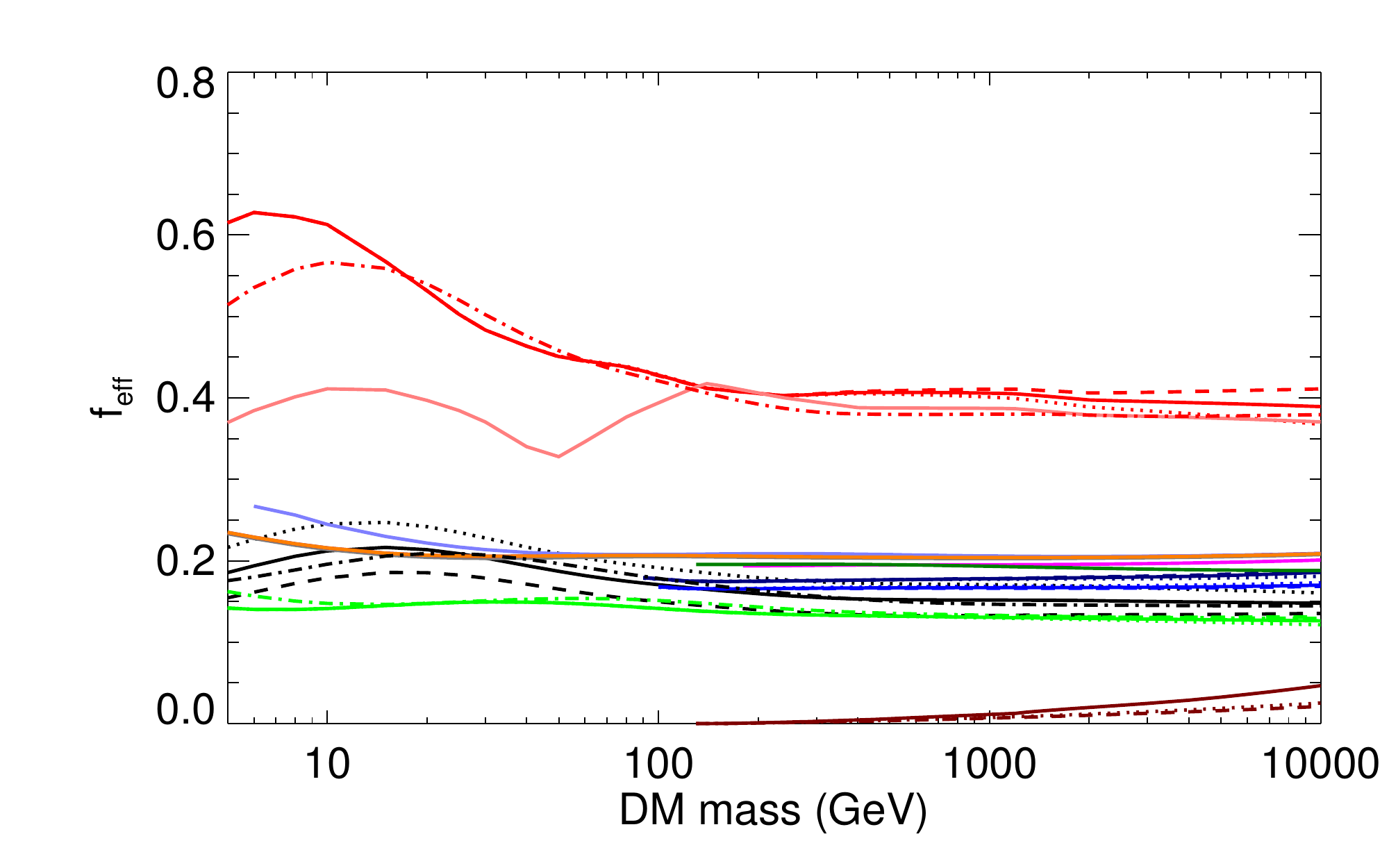}
\includegraphics[width=0.40\textwidth]{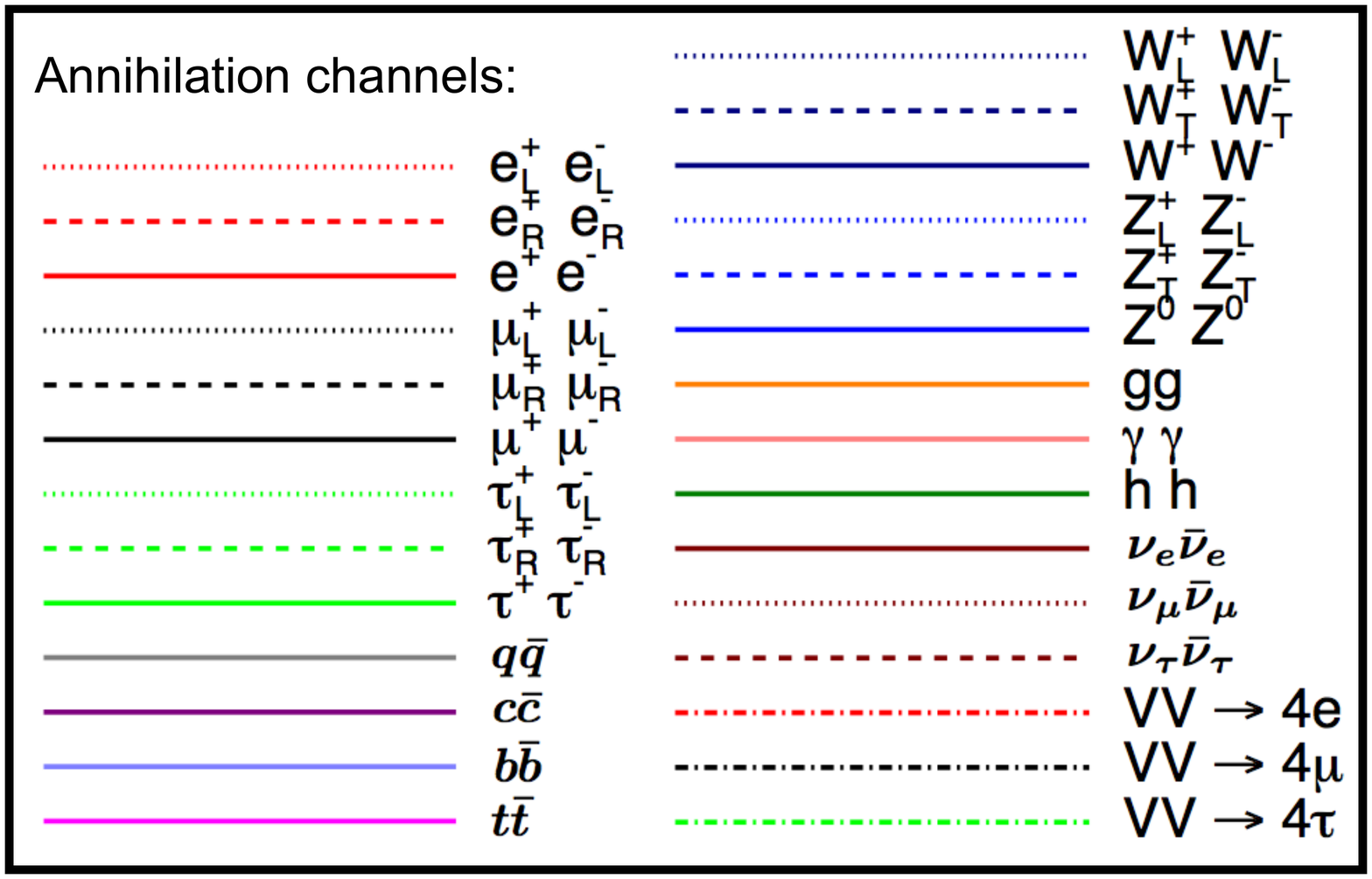}  \\
\includegraphics[width=0.45\textwidth]{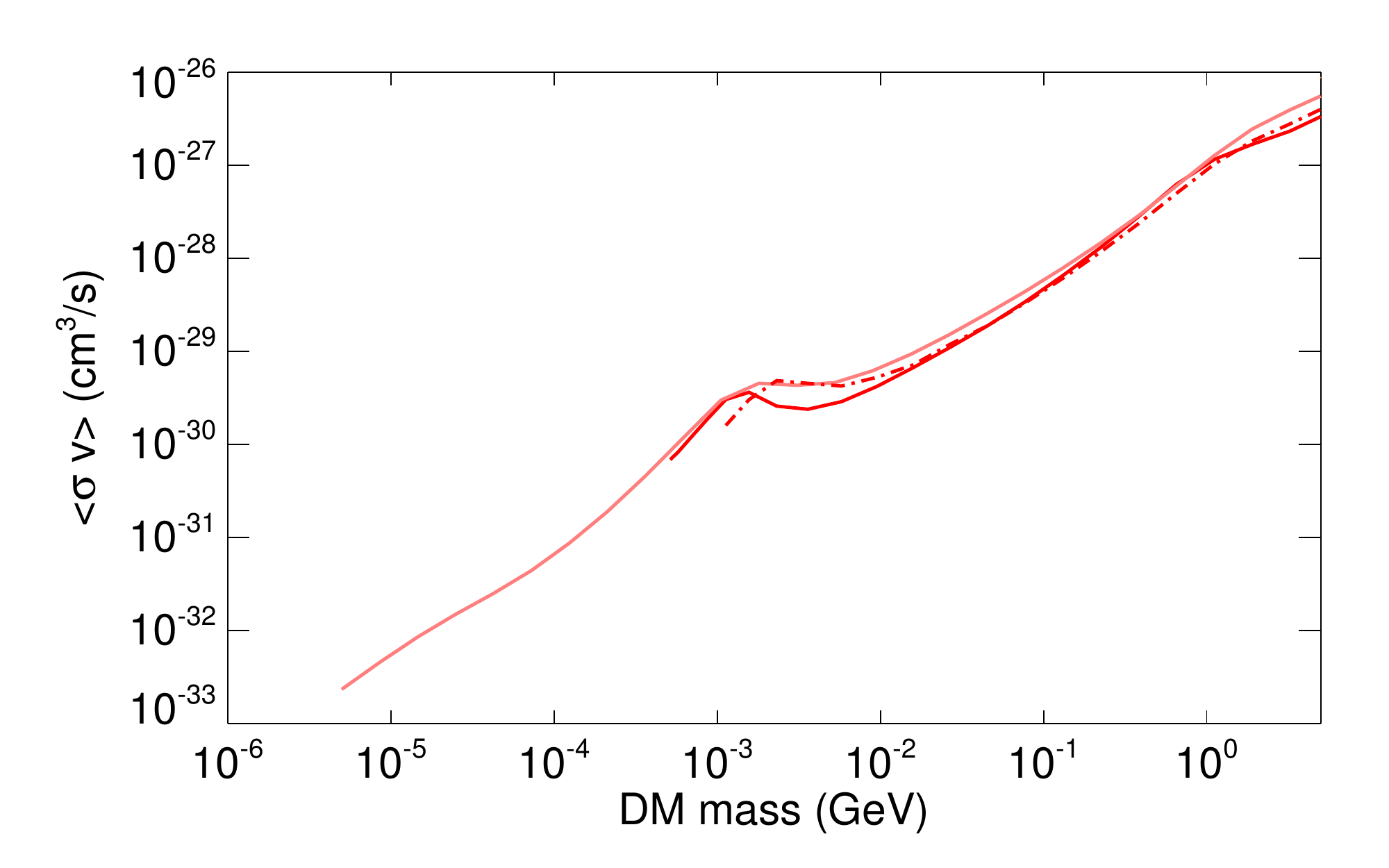}
\includegraphics[width=0.45\textwidth]{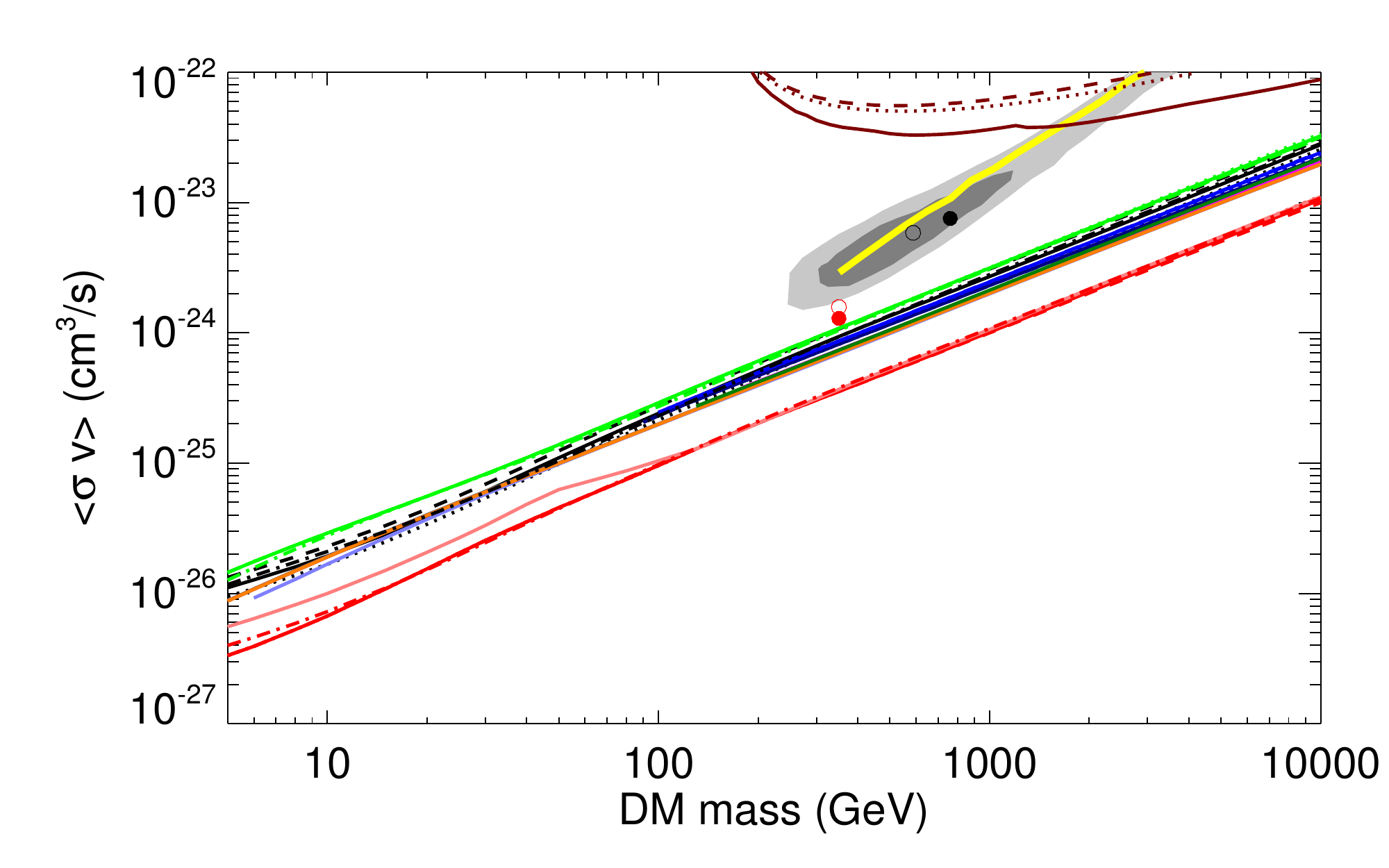}
\caption{\label{fig:dmlimits}
The upper panel shows the $f_\mathrm{eff}$ coefficients as a function of DM mass for each of a range of SM final states, as indicated in the legend. The $VV \rightarrow 4X$ states correspond to DM annihilating to a pair of new neutral vector bosons $V$, which each subsequently decay into $e^+ e^-$, $\mu^+ \mu^-$ or $\tau^+ \tau^-$ (labeled by $X$). The lower panels show the resulting estimated constraints from recent \emph{Planck} results \cite{Planck:2015xua}, as a function of DM mass, for each of the channels. The left panel covers the range from keV-scale masses up to 5 GeV, and only contains results for the $e^+ e^-$, $\gamma \gamma$ and $VV \rightarrow 4e$ channels; the right panel covers the range from 5 GeV up to 10 TeV, and covers all channels provided in the \texttt{PPPC4DMID} package \cite{Cirelli:2010xx}. The light and dark gray regions in the lower right panel correspond to the $5\sigma$ and $3\sigma$ regions in which the observed positron fraction can be explained by DM annihilation to $\mu^+ \mu^-$, for a cored DM density profile (necessary to evade $\gamma$-ray constraints), taken from \cite{Cirelli:2008pk}. The solid yellow line corresponds to the preferred cross section for the best fit 4-lepton final states identified by \cite{Boudaud:2014dta}, who argued that models in this category can still explain the positron fraction without conflicts with non-observation in other channels. The red and black circles correspond to models with $4e$ (red) and $4\mu$ (black) final states, fitted to the positron fraction in \cite{Lopez:2015uma}; as in that work, filled and open circles correspond to different cosmic-ray propagation models.}
\end{figure*}

In this section we apply the results of this work to the Standard Model final states provided in the \texttt{PPPC4DMID} package \cite{Cirelli:2010xx}, following the earlier analysis by \cite{Cline:2013fm}. \texttt{PPPC4DMID} provides the fluxes of positrons and photons produced by DM annihilation to 28 Standard Model final states, for masses ranging from 5 GeV to 100 TeV. The fluxes include electroweak final state radiation \cite{Ciafaloni:2010ti}, which can be crucial at high DM masses; thus, for example, there is a small but non-zero $f_\mathrm{eff}$ for annihilation to $\nu \bar{\nu}$, arising from the electroweak corrections.

 We only show results up to 10 TeV, as above this mass, additional processes not considered in \cite{Slatyer:2009yq} may contribute to the cooling of electrons and photons in the early universe. However, we expect that the $f_\mathrm{eff}$ results will be similar at higher DM masses, since the processes neglected in \cite{Slatyer:2009yq} will only modify the details of the pair production cascade experienced by particles injected at high energy, which occurs much faster than a Hubble time and so is anticipated to have little effect on the fraction of power eventually deposited.
 
 We take the positron and photon spectra provided in \texttt{PPPC4DMID},\footnote{\texttt{http://www.marcocirelli.net/PPPC4DMID.html}} assume that by charge symmetry the spectrum of $e^-$ is equal to that of $e^+$, and integrate the resulting $e^+ e^-$ and photon spectra over the $f_\mathrm{eff}(E)$ curves presented in Fig. \ref{fig:feff}. Specifically, if a given annihilation produces a spectrum $(dN/dE)_{e^+}$ of positrons, and a spectrum $(dN/dE)_{\gamma}$ of photons, the weighted $f_\mathrm{eff}(m_\chi)$ curve is given by:
 \begin{equation} f_\mathrm{eff}(m_\chi) = \frac{\int_0^{m_\chi} E dE \left[ 2 f^{e^+ e^-}_\mathrm{eff}(E)  \left(\frac{dN}{dE} \right)_{e^+} + f^{\gamma}_\mathrm{eff}(E)  \left(\frac{dN}{dE} \right)_{\gamma} \right]}{2 m_\chi}, \end{equation} 
 where $E$ is the energy of the positron or photon (the factor of 2 comes from the electron spectrum).
 
We use the best-estimate $f_\mathrm{eff}(E)$ curves suited for the ``3 keV'' baseline prescription, as we wish to apply the constraints derived by the \emph{Planck} Collaboration \cite{Planck:2015xua}. We take the \emph{Planck} limit to be $f_\mathrm{eff} \langle \sigma v \rangle/m_\chi < 4.1 \times 10^{-28}$ cm$^3$/s/GeV, which is the bound set by the \emph{Planck} temperature and polarization data. The limit can be strengthened by $\sim 15\%$ by the inclusion of either lensing or external datasets. We show both the $f_\mathrm{eff}(m_\chi)$ curves for each channel and the resulting constraints on the annihilation cross section in Fig. \ref{fig:dmlimits}. For DM masses below 5 GeV, we show estimated limits for DM annihilation to $e^+ e^-$, $\gamma \gamma$ or  $VV \rightarrow 4e$ -- the last channel corresponds to DM annihilating to a pair of intermediate vector bosons, each of which subsequently decays to $e^+ e^-$. The results for the first two channels are taken simply from Fig. \ref{fig:feff}, whereas for $VV \rightarrow 4e$, the $e^+ e^-$ spectrum is taken to be constant in $dN/dE$ between the kinematic limits $m_e < E < m_\chi - m_e$. For both channels producing $e^+ e^-$, there will also be an FSR contribution, but its effect is generally small (at the percent level).

In general, we see that the final states considered fall into three categories:
\begin{itemize}
\item Final states where the bulk of the power proceeds into $e^+ e^-$ and photons, where at masses above 100 GeV the constraint approaches $\langle \sigma v \rangle \lesssim 10^{-27} (m_\chi/1 \text{GeV})$ cm$^3$/s.
\item Annihilation to neutrinos, where the constraint arises entirely from electroweak corrections, and is negligible below $\sim 200$ GeV; at $\mathcal{O}$(TeV) masses, cross sections as low as a few $\times 10^{-23}$ cm$^3$/s can be constrained. Interestingly, this bound is competitive with that placed by IceCube from observations of galaxy clusters \cite{Aartsen:2013dxa}, the Galactic Center \cite{Aartsen:2015xej}, and the Milky Way halo \cite{Aartsen:2014hva}, and unlike those limits is independent of uncertainties in the local DM density, the DM distribution, and the amount of DM substructure.
\item A band with a width of roughly a factor of $150\%$ in $\langle \sigma v \rangle$ that encompasses all the other channels studied, which at high masses corresponds to $\langle \sigma v \rangle \lesssim 2-3 \times 10^{-27} (m_\chi/1 \text{GeV}) $ cm$^3$/s.
\end{itemize}

Accordingly, for any linear combination of these final states that does not contain a significant branching ratio for DM annihilation directly to neutrinos, one must have $\langle \sigma v\rangle \lesssim 3 \times 10^{-27} (m_\chi/1 \text{GeV}) $ cm$^3$/s. It is thus challenging to obtain the correct thermal relic cross section for $s$-wave annihilating DM with mass much below $m_\chi \sim 10$ GeV, without violating these limits (although models with suppressed annihilation at late times may still be viable, e.g. asymmetric DM models or the scenarios proposed in \cite{D'Agnolo:2015koa, Izaguirre:2015yja}). At higher masses, the cross sections constrained are well above the thermal relic value, but are highly relevant for DM explanations of the positron excess observed by PAMELA \cite{Adriani:2008zr}, Fermi \cite{2012PhRvL.108a1103A} and AMS-02 \cite{PhysRevLett.113.121101}. For example, \cite{Boudaud:2014dta} identified a favored scenario by which to explain this excess via DM annihilation, respecting all existing constraints: 0.5-1 TeV DM annihilating through a light mediator to a mixture of $75\%$ $\tau^+ \tau^-$ and $25\%$ $e^+e^-$. The required cross section is roughly $\langle \sigma v \rangle \sim 5 \times 10^{-24}$ cm$^3$/s at 500 GeV and roughly $\sim 1.5 \times 10^{-23}$ cm$^3$/s at 1 TeV. This is clearly above the bounds for the contributing annihilation channels in Fig. \ref{fig:dmlimits}. Similarly, \cite{Lopez:2015uma} found that only annihilation through a light mediator to muons or electrons could explain the excess without violating bounds from studies of dwarf galaxies in gamma rays. The preferred cross section was $\sim 1.3-1.6 \times 10^{-24}$ cm$^3$/s at a DM mass of $\sim$ 350 GeV for the $4e$ final state, and $\sim 6-8 \times 10^{-24}$ cm$^3$/s at a DM mass of $\sim 600-750$ GeV for the $4\mu$ final state. Again, these preferred points are well above the bounds we derive for the corresponding channels.

Fig. \ref{fig:dmlimits} shows (1) the cross section band favored for the models of \cite{Boudaud:2014dta} as a solid yellow line, (2) the points favored by \cite{Lopez:2015uma} as circles in red ($4e$ final state) or black ($4\mu$ final state), and (3) a gray contour describing the region in which DM annihilation to $\mu^+\mu^-$ can explain the AMS-02 data \cite{Cirelli:2008pk}. These favored regions should be compared, respectively, to (1) an appropriately weighted combination of the green dot-dashed and red dot-dashed lines in Fig. \ref{fig:dmlimits} (it is a conservative choice to take the green dot-dashed line to set the limit), (2) the red dot-dashed and black dot-dashed lines, and (3) the solid black line. We see that in general these models are ruled out by more than a factor of 2, and up to an order of magnitude. The models presented in \cite{Cholis:2013psa} were studied in \cite{Planck:2015xua}, and are similarly ruled out by these constraints.

\section{Conclusions}
\label{sec:conclusions}

In this work, we have mapped out the effective deposition efficiency, $f_\mathrm{eff}$, for photons and $e^+ e^-$ pairs injected into the early universe by DM annihilation, at $\mathcal{O}$(keV-TeV) kinetic energies. We have made our key numerical results available at  \texttt{http://nebel.rc.fas.harvard.edu/epsilon}. 

We have demonstrated that it is sufficient to use a simple universal weighting function to convert any redshift-dependent energy deposition history (originating from conventional $s$-wave-dominated DM annihilation) into a redshift-independent efficiency factor, confirming earlier results for the WIMP regime. This efficiency factor then converts \emph{Planck} limits on energy injection from the CMB into constraints on the DM model in question. Any DM model featuring dominantly $s$-wave annihilation, with the only redshift dependence in the annihilation rate arising from the DM density squared, can thus be constrained using these data. We emphasize that this does not subsume the limits presented by the \emph{Planck} Collaboration, but confirms the validity of the procedure employed in that analysis, and demonstrates how to extend those bounds to more general DM models.

We have performed an example calculation for 28 different annihilation channels over the DM mass range 5 GeV -- 10 TeV, and demonstrated that except for annihilation directly into neutrinos, at high masses the constraint from the CMB generically lies in the range $\langle \sigma v\rangle \lesssim 1-3 \times 10^{-27} (m_\chi/1 \text{GeV}) $ cm$^3$/s. Even for annihilation to neutrinos, at TeV-scale masses we can set an upper bound on the cross section of order a few times $10^{-23}$ cm$^3$/s, comparable to constraints from IceCube observations of the Milky Way and galaxy clusters; this bound arises from electroweak final state radiation producing $e^+e^-$ pairs and photons.

The constraints we obtain appear to rule out annihilating DM models proposed to explain the observed excess of cosmic-ray positrons, including models which could not be excluded by any previous studies. This constraint could be evaded if the DM annihilation is suppressed at low velocities or early times. 

\section*{Acknowledgements}

The author is grateful to the Mainz Institute for Theoretical Physics (MITP) and the Korea Institute for Advanced Study (KIAS) for their hospitality and support during the completion of this work, and thanks Gilly Elor, Josh Ruderman and Raffaele Tito d'Agnolo for testing these results. The author also thanks Aaron Vincent, Marco Cirelli, Bhaskar Dutta, Patrick Fox, Silvia Galli, Fabio Iocco, Hongwan Liu and Nicholas Rodd for helpful discussions. This work is supported by the U.S. Department of Energy under grant Contract Numbers DE$-$SC00012567 and DE$-$SC0013999. This research made use of the IDL Astronomy UserÕs Library at Goddard.

\onecolumngrid
\appendix

\section{Supplementary Materials}

We make available\footnote{\texttt{http://nebel.rc.fas.harvard.edu/epsilon}} a selection of \texttt{.fits} and \texttt{.dat} files containing the $f_\mathrm{eff}$ values presented in Figs. \ref{fig:feff} and \ref{fig:dmlimits} (with the exception of the $VV \rightarrow 4e$ channel below 5 GeV, which can be trivially obtained -- at least at the level we have calculated it -- from the results for $e^+ e^-$), as well as the universal annihilation weighting function shown in Fig. \ref{fig:weightingfn}. We also provide an example \texttt{Mathematica} notebook to demonstrate the use of the \texttt{.fits} files. 

The \texttt{.fits} files are as follows:

\vskip 5mm

{\bf  \texttt{feff\_summary\_species\_base.fits}}

Here ``species'' can be ``elec'' (representing $e^+ e^-$ pairs) or ``phot'' (representing photons), and ``base'' can be ``SSCK'' or ``3 keV'', corresponding to the choice of baseline ionization prescription. For example, the ``3 keV'' files should be used with the constraints on DM annihilation of \cite{Planck:2015xua}. Each file contains the following arrays:
\begin{itemize} 
\item \texttt{REDSHIFT}: this 63-element array provides the abscissa for deposition redshift.
\item \texttt{WEIGHTFN}: this 63-element array provides the weighting function shown in the third panel of Fig. \ref{fig:weightingfn}, sampled at the redshifts given by the \texttt{REDSHIFT} array, and normalized so that $\int W(z) d\ln(1+z) = 1$ (up to numerical error).
\item \texttt{LOG10ENERGY:} this 40-element array gives $\log_{10}(E/\mathrm{eV})$, where $E$ is the kinetic energy of one member of the pair of injected particles (for $e^+ e^-$ pairs) or the energy of a single injected photon.
\item \texttt{FEFF\_BEST:} this $40$-element array gives $f_\mathrm{eff}$ computed using the appropriate $f^\mathrm{base}(z)$ curve, weighted by the weighting function.
\item \texttt{FEFF\_RESCALING:} this $40$-element array gives $f_\mathrm{eff}$ computed using the appropriate $f^\mathrm{sim}(z)$ curve, weighted by the weighting function.
\item \texttt{FEFF\_BEST\_600:} this $40$-element array gives $f_\mathrm{eff}$ computed using the appropriate $f^\mathrm{base}(z)$ curve, evaluated at $z=600$.
\end{itemize}

\vskip 5mm

{\bf  \texttt{feff\_PPPC\_3keV.fits}}

This file contains the $f_\mathrm{eff}$ curves for different DM masses and annihilation channels as shown in Fig. \ref{fig:dmlimits}, at DM masses above 5 GeV. Below 5 GeV, the results for the $e^+ e^-$, $\gamma \gamma$ and $VV \rightarrow 4e$ channels are easily obtained from the \texttt{feff\_summary\_species\_3keV.fits} files. The arrays in this file are:
\begin{itemize} 
\item \texttt{DM\_MASS:} this 62-element array gives the DM mass in GeV.
\item \texttt{CHANNEL:} this 28-element array labels the channels as shown in the legend of Fig. \ref{fig:dmlimits}.
\item \texttt{FEFF:} this $28 \times 62$ array contains the $f_\mathrm{eff}$ values for each DM mass and channel, derived using the appropriate $f^\mathrm{3keV}(z)$ curves and the weighting function.
\end{itemize}

The \texttt{.dat} files simply contain condensed summaries of the \texttt{.fits} files, and are as follows:

\vskip 5mm

{\bf  \texttt{feff\_PPPC\_3keV.dat}}

As \texttt{feff\_PPPC\_3keV.fits}, except that the $f_\mathrm{eff}$ values are laid out in a grid, with columns corresponding to different channels (first row describes the different final states) and rows corresponding to different DM masses (first column gives the DM mass in GeV).

\vskip 5mm

{\bf  \texttt{feff\_base.dat}}

Contains a subset of the information in \texttt{feff\_summary\_electron\_base.fits} and \texttt{feff\_summary\_photon\_base.fits}: the first column is the \texttt{LOG10ENERGY} array from these files, describing the injection energy; the second is the best-estimate $f_\mathrm{eff}$ value (computed using the $f^\mathrm{base}(z)$ curves and the weighting function) for $e^+ e^-$ pairs; the third is the best-estimate $f_\mathrm{eff}$ for photons. ``base'' can take the values ``SSCK'' or ``3 keV'' and should be chosen based on the ionization prescription used in the analysis to which comparison is desired.

\vskip 5mm

{\bf  \texttt{weighting\_function.dat}}

Contains the weighting function $W(z)$ shown in the third panel of Fig. \ref{fig:weightingfn}, sampled at the redshifts $1+z$ given in the first column.

\section{Review of Principal Component Analysis}
\label{app:pca}

Suppose we have some arbitrary energy deposition history, characterized by a (corrected) efficiency function $f(z)$. Let us discretize the $f(z)$ function into $n_z$ bins centered at $z_i$, $i=1..n_z$, thus describing the energy deposition by a column vector $\vec{v} = (f(z_1), f(z_2), f(z_3), ..., f(z_{n_z}))$. We follow \cite{Finkbeiner:2011dx} in treating the energy deposition profile $f(z)$ in each bin as a Gaussian in $\ln(1+z)$, normalized such that its integral with respect to $d\ln(1+z)$ is equal to the logarithmic bin width $\Delta \ln(1+z)$.

Within the approximation of linearity (which is a good approximation for energy injections at the level constrained by \emph{Planck} \cite{Finkbeiner:2011dx}), the impact of this deposition history on the anisotropies of the cosmic microwave background can be written in the form,
\begin{equation} h \equiv \Delta C_\ell = T \vec{v}, \end{equation}
where $T$ is some transfer matrix of dimension $n_\ell \times n_z$, and $n_\ell$ is the number of multipoles we consider. Following \cite{Finkbeiner:2011dx}, in our notation each element of $T$ ($T_{\ell i}$) is itself a three-element vector, holding the perturbations to the TT,  TE and EE anisotropy spectra at that $\ell$. $T$ can be computed element-by-element, as described in \cite{Finkbeiner:2011dx}, by taking $\vec{v}_j$ such that $v_i = \delta_{ij}$, and computing the effects on the CMB using \texttt{RECFAST} \cite{Seager:1999} (or a similar code such as \texttt{CosmoRec} \cite{Chluba:2010ca} or \texttt{HyRec} \cite{AliHaimoud:2010dx}) and \texttt{CAMB} \cite{2000ApJ...538..473L}. Similarly, we can determine a corresponding $n_\ell \times 6$ transfer matrix describing the effect of varying the standard six cosmological parameters.

Once $T$ has been mapped out, we can construct the $n_z \times n_z$ Fisher matrix $F_e$ as,
\begin{equation} (F_e)_{ij} = \sum_\ell T_{\ell i}^T \cdot \Sigma_\ell^{-1} T_{\ell j}, \label{eq:fisher} \end{equation}
where $\Sigma_\ell$ is the appropriate covariance matrix for the anisotropy spectra (see \cite{Finkbeiner:2011dx} for the explicit form). Schematically, $F_e = T^T \Sigma^{-1} T$; it is a symmetric matrix, due to the symmetry of $\Sigma$, and its diagonal elements $(F_e)_{ii}$ describe the (squared) signal significance per energy deposition at redshift $z_i$, before marginalization over the existing cosmological parameters. 

To perform this marginalization, one first constructs an expanded Fisher matrix $F_0$, using Eq. \ref{eq:fisher} but expanding $T$ to include the effect of the standard six cosmological parameters. We can write:
\begin{equation} F_0 = \left(\begin{matrix}F_e & F_v \\ F_v^T & F_c \end{matrix} \right),\end{equation}
where $F_e$ is the pre-marginalization Fisher matrix as defined above, $F_c$ is the Fisher matrix for the cosmological parameters only, and $F_v$ describes the cross terms. Then the \emph{marginalized} Fisher matrix, describing the detectability of an energy injection after projecting out the perturbations parallel to the standard cosmological parameters, is given by $F_z = F_e - F_v F_c^{-1} F_v^T$ (see \cite{Finkbeiner:2011dx} for a discussion).

The principal components are given by the eigenvectors of $F_z$, which we will denote $\{ \vec{P}_i \}$, $i=1..n_z$. By construction, they form an orthogonal basis for arbitrary $f(z)$ histories, and we can choose them to be orthonormal; as such, one can write the vector corresponding to any arbitrary $f(z)$ in the form $\vec{v} = \sum \alpha_i \vec{P}_i$, where $\vec{P}_i$ is the $i$th principal component and $\alpha_i = \vec{v} \cdot \vec{P}_i$. Each principal component perturbs the CMB anisotropies with significance proportional to the square root of the corresponding eigenvalue (to the degree that the approximations inherent in the Fisher matrix analysis are valid). Consequently, if the principal components are ranked in order of their eigenvalues, the detectability of an arbitrary energy deposition history can be described approximately by $\alpha_1 = \vec{v} \cdot \vec{P}_1$, the dot product with the first principal component. In this sense, $\vec{P}_1$ acts as a redshift-dependent weighting function that determines the detectability of an arbitrary energy deposition history. The accuracy of this ``weighting function'' approximation will be controlled by the fraction of the total variance described by the first eigenvalue (it may also break down for specific energy deposition histories where $\vec{v}$ is accidentally nearly orthogonal to $\vec{P}_1$).

\section{Deriving the Universal Weighting Function for Annihilating DM}
\label{app:weightfn}

Rather than working in terms of energy deposition histories localized in redshift, one can perform a similar PCA starting with any sample of $f(z)$ functions (see \cite{Finkbeiner:2011dx} for a more complete discussion); the resulting principal components will be applicable to any energy deposition history lying within the space spanned by these basis functions. For the purposes of setting CMB constraints, any conventional DM annihilation model can be described as a linear combination of photons and $e^+ e^-$ pairs injected over a range of energies. 

Consequently, we can perform a PCA applicable to all conventional DM annihilation models, by taking as our basis functions the $f(z)$ curves plotted in Fig. \ref{fig:fDM}. We choose 20 log-spaced injection energies for photons and $e^+ e^-$ pairs, thus giving rise to 40 $f(z)$ curves as our sample. As previously, we can sample these curves at $n_z$ redshifts, translating them into $n_z$-element column vectors, which we will denote $\vec{u}_i$ for $i=1..40$. (More generally, we will denote the number of models included in the PCA as $n_M$: in this case $n_M = 40$.) Let $U$ be the $n_z \times n_M$ matrix with the $i$th column given by $\vec{u}_i$. A general DM annihilation model can be expressed as $\vec{v} = U \vec{\alpha}$, where $\vec{\alpha}$ now describes the coefficients of the (\emph{not} orthogonal) basis vectors $\vec{u}_i$. Thus we can describe the matrix $U$ as mapping from $n_M$-dimensional ``coefficient space'' into $n_z$-dimensional ``redshift space''.

Following the same procedure as previously, we can compute the impact on the CMB of each of the $\vec{u}_i$ histories, thus constructing a $n_\ell \times n_M$ transfer matrix $T_M$ that maps from ``coefficient space'' to the anisotropies of the CMB. In terms of the earlier transfer matrix $T$, $T_M = T U$. Let us label the associated $n_M \times n_M$ Fisher matrix (after marginalization over the standard cosmological parameters) as $F_M$, and the expanded pre-marginalization Fisher matrix as $F_{0M}$.

Since the transfer matrix for the cosmological parameters is unchanged, writing $T_M$ in terms of $T$ yields the following relation between $F_0$ and $F_{0M}$:
\begin{equation}F_{0M} = \left( \begin{array}{cc} U^T F_{e} U & U^T F_{v} \\ (U^T F_{v})^T & F_{c} \end{array} \right). \end{equation}
Consequently, the marginalized Fisher matrix becomes,
\begin{equation} F_M = U^T F_{e} U - U^T F_{v} F_{c}^{-1} F_{v}^T U = U^T F_z U.\end{equation}

Diagonalizing $F_M$ now yields a set of principal components specialized to the case of DM annihilation; these principal components are $n_M$-element vectors, and describe coefficient vectors $\vec{\alpha}$ for the original basis $f(z)$ curves. Accordingly, they are orthonormal in ``coefficient space'', \emph{not} ``redshift space''. If we denote these principal components as $\{\vec{p}_i\}$, $i=1..n_M$, then following the reasoning above, we can write $\vec{\alpha} = \sum_i (\vec{\alpha} \cdot \vec{p}_i) \vec{p}_i$, and the detectability of the energy deposition history characterized by the coefficient vector $\vec{\alpha}$ is determined approximately by $\vec{\alpha} \cdot \vec{p}_1$. 

Due to the similarity of the $f(z)$ curves for different energies and different species, when the eigenvalues are calculated, the first principal component accounts for more than $99\%$ of the variance. Accordingly, having specialized to the case of DM annihilation, the ``weighting function'' approximation is expected to be accurate at the sub-percent level.

However, we would prefer to express this dot product directly in terms of the energy deposition history $f(z)$, or its discretized counterpart $\vec{v}$, rather than its coefficients with respect to a somewhat arbitrary basis. To achieve this, note that since $\vec{v} = U \vec{\alpha}$, we can write,
\begin{align} U^T F_z \vec{v} = U^T F_z U \vec{\alpha} = F_M   \sum_{i=1}^{m_N} (\vec{\alpha} \cdot \vec{p_i}) \vec{p_i} = \sum_{i=1}^{m_N} \lambda_i (\vec{\alpha} \cdot \vec{p_i}) \vec{p_i} \end{align}
Here $\lambda_i$ is the $i$th eigenvalue of $F_M$, and so $F_M \vec{p}_i = \lambda_i \vec{p}_i$ by definition. Taking the dot product with $\vec{p}_1$ then yields,
\begin{align} p_1 \cdot (U^T F_z \vec{v}) =  \lambda_1 (\vec{\alpha} \cdot \vec{p_1}), \end{align}
and consequently,
\begin{align} \vec{\alpha} \cdot \vec{p_1} = \frac{1}{\lambda_1}  p_1^T U^T F_z \vec{v} = \left( \frac{(U \vec{p}_1)^T F_z}{\lambda_1} \right) \cdot \vec{v}.\end{align}

If a constraint on \emph{some} deposition history $f_0(z)$ (lying within the space spanned by conventional DM annihilation) has been computed, therefore, the constraint on any \emph{other} $f(z)$ (at the same DM mass and cross section) can be computed by rescaling the bound by a factor:
\begin{align} f_\mathrm{eff} =  \frac{\left( \frac{(U \vec{p}_1)^T F_z}{\lambda_1} \right) \cdot \vec{v}}{\left( \frac{(U \vec{p}_1)^T F_z}{\lambda_1} \right) \cdot \vec{v}_0} = \left[ \frac{(U \vec{p}_1)^T F_z }{\left( (U \vec{p}_1)^T F_z \right) \cdot \vec{v}_0} \right]  \cdot \vec{v}  \end{align}
A higher $f_\mathrm{eff}$ corresponds to a more detectable $f(z)$, and hence to a stronger bound. The quantity in square brackets now acts as a weighting vector $\vec{W}$ which converts $\vec{v}$ to $f_\mathrm{eff}$. In principle, we could use any vector $\vec{v}_0$ to determine the reference bound, but in practice the usual choice is $f(z)=1$, so $f_\mathrm{eff}$ represents ``effective deposition efficiency'' relative to the case corresponding to total prompt energy deposition (with one of the simplified ionization prescriptions). This choice yields:
\begin{equation} \vec{W} =  \frac{(U \vec{p}_1)^T F_z }{\sum_i \left( (U \vec{p}_1)^T F_z \right)_i} . \end{equation}

We can define a \emph{continuous} weighting function $W(z)$ by interpolation, fixing:
\begin{equation} W(z_i) = \frac{(\vec{W})_i}{d\ln(1+z)}. \end{equation}
Then the normalization condition $\sum_i (\vec{W})_i = 1$ translates into $\sum_i W(z_i) d\ln(1+z) = 1$, and so when the continuous limit is taken, $\int W(z) d\ln(1+z) = 1$. We can likewise rewrite $f_\mathrm{eff}$ in terms of the continuous functions $f(z)$ and $W(z)$,
\begin{align} f_\mathrm{eff} & = \sum_i f(z_i) (\vec{W})_i = \sum_i f(z_i) W(z_i) d\ln(1+z) \nonumber \\
& \rightarrow \int f(z) W(z) d\ln(1+z). \end{align}

This is the weighting function provided in the main text.

\section{Expected Change to the Weighting Function Due to Ionization Prescription}
\label{app:ionpca}

As discussed in Sec. \ref{sec:cmbbounds}, the weighting function has some dependence on the presumed baseline ionization prescription, even though the $f^\mathrm{SSCK}(z)$ and $f^\mathrm{3keV}(z)$ curves correspond to identical physical ionization histories (when properly combined with the appropriate ionization prescriptions). There is also a \emph{separate} dependence on the presumed baseline ionization prescription when the $f^\mathrm{sim}(z)$ curve is used, but in this case the dependence is physical; it affects the mapping from DM models into the CMB, by changing the transfer matrix $T_M$ (and $T$).

In the notation of Appendix \ref{app:weightfn}, when $f^\mathrm{SSCK}(z)$ or $f^\mathrm{3keV}(z)$ curves are used, the $F_M$ Fisher matrix (and resulting principal components) are unaffected by the choice of ionization prescription, as these depend only on the (physical) CMB signatures of particular DM annihilation models. However, the $U$ and $F_z$ matrices, which involve mappings between DM models and $f(z)$ curves, or $f(z)$ curves and the CMB, \emph{are} affected.

Let $U_1$ and $U_2$ denote the matrices that hold the $f(z)$ curves for two different baseline ionization prescriptions, and let $i_1(z)$, $i_2(z)$ describe the fraction of deposited power proceeding into ionization (as a function of redshift) in those two prescriptions. Let $I_1$ and $I_2$ define the $n_z \times n_z$ diagonal matrices such that $(I_a)_{jk} = i_a(z_j) \delta_{jk}$. Then since by the definition of $f^\mathrm{base}(z)$ curves, the power into ionization at any redshift is invariant, we can write $I_1 U_1 = I_2 U_2$.

Now if $F_{z1}$ and $F_{z2}$ are the Fisher matrices corresponding to the two baseline ionization prescriptions, we can write $F_M = U_1^T F_{z1} U_1 = U_1^T I_1 I_1^{-1} F_{z1} I_1^{-1} I_1 U_1 = U_2^T I_2 I_1^{-1} F_{z1} I_1^{-1} I_2 U_2 = U_2^T F_{z2} U_2$. It follows that,
\begin{equation} F_{z2} =  I_2 I_1^{-1} F_{z1} I_1^{-1} I_2.\end{equation}
Then we can immediately write down the relationship between the weighting vectors corresponding to the two prescriptions,
\begin{align} \vec{W}_2 & =  \frac{(U_2 \vec{p}_1)^T F_{z2} }{\sum_i \left( (U_2 \vec{p}_1)^T F_{z2} \right)_i} \nonumber  \\
& = \frac{(I_2^{-1} I_1 U_1 \vec{p}_1)^T I_2 I_1^{-1} F_{z1} I_1^{-1} I_2 }{\sum_i \left( (I_2^{-1} I_1 U_1 \vec{p}_1)^T I_2 I_1^{-1} F_{z1} I_1^{-1} I_2\right)_i} \nonumber \\
& =  \frac{(U_1 \vec{p}_1)^T F_{z1} I_1^{-1} I_2 }{\sum_i \left( (U_1 \vec{p}_1)^T  F_{z1} I_1^{-1} I_2\right)_i} \nonumber \\
& = \vec{W}_1 I_1^{-1} I_2 \frac{\sum_i \left( (U_1 \vec{p}_1)^T  F_{z1}\right)_i}{\sum_i \left( (U_1 \vec{p}_1)^T  F_{z1} I_1^{-1} I_2\right)_i } . \end{align}

We see that if $I_1$ and $I_2$ only vary by a redshift-independent normalization factor, the effect on $\vec{W}$ cancels out entirely, and this is a quite a good approximation at all redshifts of interest for the signal (see e.g. Paper II \cite{inprep}). Thus while the redshift-dependence of $I_1^{-1} I_2$ \emph{does} slightly distort the shape of the weighting function, the impact on the extracted $f_\mathrm{eff}$ is negligible (as demonstrated explicitly in Sec. \ref{sec:cmbbounds}).



\bibliography{deposition}

\begin{thebibliography}{53}
\expandafter\ifx\csname natexlab\endcsname\relax\def\natexlab#1{#1}\fi
\expandafter\ifx\csname bibnamefont\endcsname\relax
  \def\bibnamefont#1{#1}\fi
\expandafter\ifx\csname bibfnamefont\endcsname\relax
  \def\bibfnamefont#1{#1}\fi
\expandafter\ifx\csname citenamefont\endcsname\relax
  \def\citenamefont#1{#1}\fi
\expandafter\ifx\csname url\endcsname\relax
  \def\url#1{\texttt{#1}}\fi
\expandafter\ifx\csname urlprefix\endcsname\relax\def\urlprefix{URL }\fi
\providecommand{\bibinfo}[2]{#2}
\providecommand{\eprint}[2][]{\url{#2}}

\bibitem[{\citenamefont{Adams et~al.}(1998)\citenamefont{Adams, Sarkar, and
  Sciama}}]{Adams:1998nr}
\bibinfo{author}{\bibfnamefont{J.~A.} \bibnamefont{Adams}},
  \bibinfo{author}{\bibfnamefont{S.}~\bibnamefont{Sarkar}}, \bibnamefont{and}
  \bibinfo{author}{\bibfnamefont{D.}~\bibnamefont{Sciama}},
  \bibinfo{journal}{Mon.Not.Roy.Astron.Soc.} \textbf{\bibinfo{volume}{301}},
  \bibinfo{pages}{210} (\bibinfo{year}{1998}), \eprint{astro-ph/9805108}.

\bibitem[{\citenamefont{Chen and Kamionkowski}(2004)}]{Chen:2003gz}
\bibinfo{author}{\bibfnamefont{X.-L.} \bibnamefont{Chen}} \bibnamefont{and}
  \bibinfo{author}{\bibfnamefont{M.}~\bibnamefont{Kamionkowski}},
  \bibinfo{journal}{Phys. Rev.} \textbf{\bibinfo{volume}{D70}},
  \bibinfo{pages}{043502} (\bibinfo{year}{2004}), \eprint{astro-ph/0310473}.

\bibitem[{\citenamefont{Padmanabhan and Finkbeiner}(2005)}]{Padmanabhan:2005es}
\bibinfo{author}{\bibfnamefont{N.}~\bibnamefont{Padmanabhan}} \bibnamefont{and}
  \bibinfo{author}{\bibfnamefont{D.~P.} \bibnamefont{Finkbeiner}},
  \bibinfo{journal}{Phys. Rev.} \textbf{\bibinfo{volume}{D72}},
  \bibinfo{pages}{023508} (\bibinfo{year}{2005}), \eprint{astro-ph/0503486}.

\bibitem[{\citenamefont{{Hinshaw} et~al.}(2013)\citenamefont{{Hinshaw},
  {Larson}, {Komatsu}, {Spergel}, {Bennett}, {Dunkley}, {Nolta}, {Halpern},
  {Hill}, {Odegard} et~al.}}]{2013ApJS..208...19H}
\bibinfo{author}{\bibfnamefont{G.}~\bibnamefont{{Hinshaw}}},
  \bibinfo{author}{\bibfnamefont{D.}~\bibnamefont{{Larson}}},
  \bibinfo{author}{\bibfnamefont{E.}~\bibnamefont{{Komatsu}}},
  \bibinfo{author}{\bibfnamefont{D.~N.} \bibnamefont{{Spergel}}},
  \bibinfo{author}{\bibfnamefont{C.~L.} \bibnamefont{{Bennett}}},
  \bibinfo{author}{\bibfnamefont{J.}~\bibnamefont{{Dunkley}}},
  \bibinfo{author}{\bibfnamefont{M.~R.} \bibnamefont{{Nolta}}},
  \bibinfo{author}{\bibfnamefont{M.}~\bibnamefont{{Halpern}}},
  \bibinfo{author}{\bibfnamefont{R.~S.} \bibnamefont{{Hill}}},
  \bibinfo{author}{\bibfnamefont{N.}~\bibnamefont{{Odegard}}},
  \bibnamefont{et~al.}, \bibinfo{journal}{\apjs}
  \textbf{\bibinfo{volume}{208}}, \bibinfo{eid}{19} (\bibinfo{year}{2013}),
  \eprint{1212.5226}.

\bibitem[{\citenamefont{{Story} et~al.}(2013)\citenamefont{{Story},
  {Reichardt}, {Hou}, {Keisler}, {Aird}, {Benson}, {Bleem}, {Carlstrom},
  {Chang}, {Cho} et~al.}}]{2013ApJ...779...86S}
\bibinfo{author}{\bibfnamefont{K.~T.} \bibnamefont{{Story}}},
  \bibinfo{author}{\bibfnamefont{C.~L.} \bibnamefont{{Reichardt}}},
  \bibinfo{author}{\bibfnamefont{Z.}~\bibnamefont{{Hou}}},
  \bibinfo{author}{\bibfnamefont{R.}~\bibnamefont{{Keisler}}},
  \bibinfo{author}{\bibfnamefont{K.~A.} \bibnamefont{{Aird}}},
  \bibinfo{author}{\bibfnamefont{B.~A.} \bibnamefont{{Benson}}},
  \bibinfo{author}{\bibfnamefont{L.~E.} \bibnamefont{{Bleem}}},
  \bibinfo{author}{\bibfnamefont{J.~E.} \bibnamefont{{Carlstrom}}},
  \bibinfo{author}{\bibfnamefont{C.~L.} \bibnamefont{{Chang}}},
  \bibinfo{author}{\bibfnamefont{H.-M.} \bibnamefont{{Cho}}},
  \bibnamefont{et~al.}, \bibinfo{journal}{\apj} \textbf{\bibinfo{volume}{779}},
  \bibinfo{eid}{86} (\bibinfo{year}{2013}), \eprint{1210.7231}.

\bibitem[{\citenamefont{{Hou} et~al.}(2014)\citenamefont{{Hou}, {Reichardt},
  {Story}, {Follin}, {Keisler}, {Aird}, {Benson}, {Bleem}, {Carlstrom}, {Chang}
  et~al.}}]{2014ApJ...782...74H}
\bibinfo{author}{\bibfnamefont{Z.}~\bibnamefont{{Hou}}},
  \bibinfo{author}{\bibfnamefont{C.~L.} \bibnamefont{{Reichardt}}},
  \bibinfo{author}{\bibfnamefont{K.~T.} \bibnamefont{{Story}}},
  \bibinfo{author}{\bibfnamefont{B.}~\bibnamefont{{Follin}}},
  \bibinfo{author}{\bibfnamefont{R.}~\bibnamefont{{Keisler}}},
  \bibinfo{author}{\bibfnamefont{K.~A.} \bibnamefont{{Aird}}},
  \bibinfo{author}{\bibfnamefont{B.~A.} \bibnamefont{{Benson}}},
  \bibinfo{author}{\bibfnamefont{L.~E.} \bibnamefont{{Bleem}}},
  \bibinfo{author}{\bibfnamefont{J.~E.} \bibnamefont{{Carlstrom}}},
  \bibinfo{author}{\bibfnamefont{C.~L.} \bibnamefont{{Chang}}},
  \bibnamefont{et~al.}, \bibinfo{journal}{\apj} \textbf{\bibinfo{volume}{782}},
  \bibinfo{eid}{74} (\bibinfo{year}{2014}), \eprint{1212.6267}.

\bibitem[{\citenamefont{Sievers et~al.}(2013)}]{Sievers:2013ica}
\bibinfo{author}{\bibfnamefont{J.~L.} \bibnamefont{Sievers}}
  \bibnamefont{et~al.} (\bibinfo{collaboration}{Atacama Cosmology Telescope}),
  \bibinfo{journal}{JCAP} \textbf{\bibinfo{volume}{1310}}, \bibinfo{pages}{060}
  (\bibinfo{year}{2013}), \eprint{1301.0824}.

\bibitem[{\citenamefont{Ade et~al.}(2015)}]{Planck:2015xua}
\bibinfo{author}{\bibfnamefont{P.}~\bibnamefont{Ade}} \bibnamefont{et~al.}
  (\bibinfo{collaboration}{Planck}) (\bibinfo{year}{2015}),
  \eprint{1502.01589}.

\bibitem[{\citenamefont{Zhang et~al.}(2007)\citenamefont{Zhang, Chen,
  Kamionkowski, Si, and Zheng}}]{Zhang:2007zzh}
\bibinfo{author}{\bibfnamefont{L.}~\bibnamefont{Zhang}},
  \bibinfo{author}{\bibfnamefont{X.}~\bibnamefont{Chen}},
  \bibinfo{author}{\bibfnamefont{M.}~\bibnamefont{Kamionkowski}},
  \bibinfo{author}{\bibfnamefont{Z.-g.} \bibnamefont{Si}}, \bibnamefont{and}
  \bibinfo{author}{\bibfnamefont{Z.}~\bibnamefont{Zheng}},
  \bibinfo{journal}{Phys. Rev.} \textbf{\bibinfo{volume}{D76}},
  \bibinfo{pages}{061301} (\bibinfo{year}{2007}), \eprint{0704.2444}.

\bibitem[{\citenamefont{Galli et~al.}(2009)\citenamefont{Galli, Iocco, Bertone,
  and Melchiorri}}]{Galli:2009zc}
\bibinfo{author}{\bibfnamefont{S.}~\bibnamefont{Galli}},
  \bibinfo{author}{\bibfnamefont{F.}~\bibnamefont{Iocco}},
  \bibinfo{author}{\bibfnamefont{G.}~\bibnamefont{Bertone}}, \bibnamefont{and}
  \bibinfo{author}{\bibfnamefont{A.}~\bibnamefont{Melchiorri}},
  \bibinfo{journal}{Phys. Rev.} \textbf{\bibinfo{volume}{D80}},
  \bibinfo{pages}{023505} (\bibinfo{year}{2009}), \eprint{0905.0003}.

\bibitem[{\citenamefont{Slatyer et~al.}(2009)\citenamefont{Slatyer,
  Padmanabhan, and Finkbeiner}}]{Slatyer:2009yq}
\bibinfo{author}{\bibfnamefont{T.~R.} \bibnamefont{Slatyer}},
  \bibinfo{author}{\bibfnamefont{N.}~\bibnamefont{Padmanabhan}},
  \bibnamefont{and} \bibinfo{author}{\bibfnamefont{D.~P.}
  \bibnamefont{Finkbeiner}}, \bibinfo{journal}{Phys. Rev.}
  \textbf{\bibinfo{volume}{D80}}, \bibinfo{pages}{043526}
  (\bibinfo{year}{2009}), \eprint{0906.1197}.

\bibitem[{\citenamefont{Kanzaki et~al.}(2010)\citenamefont{Kanzaki, Kawasaki,
  and Nakayama}}]{Kanzaki:2009hf}
\bibinfo{author}{\bibfnamefont{T.}~\bibnamefont{Kanzaki}},
  \bibinfo{author}{\bibfnamefont{M.}~\bibnamefont{Kawasaki}}, \bibnamefont{and}
  \bibinfo{author}{\bibfnamefont{K.}~\bibnamefont{Nakayama}},
  \bibinfo{journal}{Prog.Theor.Phys.} \textbf{\bibinfo{volume}{123}},
  \bibinfo{pages}{853} (\bibinfo{year}{2010}), \eprint{0907.3985}.

\bibitem[{\citenamefont{Hisano et~al.}(2011)\citenamefont{Hisano, Kawasaki,
  Kohri, Moroi, Nakayama et~al.}}]{Hisano:2011dc}
\bibinfo{author}{\bibfnamefont{J.}~\bibnamefont{Hisano}},
  \bibinfo{author}{\bibfnamefont{M.}~\bibnamefont{Kawasaki}},
  \bibinfo{author}{\bibfnamefont{K.}~\bibnamefont{Kohri}},
  \bibinfo{author}{\bibfnamefont{T.}~\bibnamefont{Moroi}},
  \bibinfo{author}{\bibfnamefont{K.}~\bibnamefont{Nakayama}},
  \bibnamefont{et~al.}, \bibinfo{journal}{Phys.Rev.}
  \textbf{\bibinfo{volume}{D83}}, \bibinfo{pages}{123511}
  (\bibinfo{year}{2011}), \eprint{1102.4658}.

\bibitem[{\citenamefont{Hutsi et~al.}(2011)\citenamefont{Hutsi, Chluba, Hektor,
  and Raidal}}]{Hutsi:2011vx}
\bibinfo{author}{\bibfnamefont{G.}~\bibnamefont{Hutsi}},
  \bibinfo{author}{\bibfnamefont{J.}~\bibnamefont{Chluba}},
  \bibinfo{author}{\bibfnamefont{A.}~\bibnamefont{Hektor}}, \bibnamefont{and}
  \bibinfo{author}{\bibfnamefont{M.}~\bibnamefont{Raidal}},
  \bibinfo{journal}{\aap} \textbf{\bibinfo{volume}{535}}, \bibinfo{eid}{A26}
  (\bibinfo{year}{2011}), \eprint{1103.2766}.

\bibitem[{\citenamefont{{Galli} et~al.}(2011)\citenamefont{{Galli}, {Iocco},
  {Bertone}, and {Melchiorri}}}]{2011PhRvD..84b7302G}
\bibinfo{author}{\bibfnamefont{S.}~\bibnamefont{{Galli}}},
  \bibinfo{author}{\bibfnamefont{F.}~\bibnamefont{{Iocco}}},
  \bibinfo{author}{\bibfnamefont{G.}~\bibnamefont{{Bertone}}},
  \bibnamefont{and}
  \bibinfo{author}{\bibfnamefont{A.}~\bibnamefont{{Melchiorri}}},
  \bibinfo{journal}{\prd} \textbf{\bibinfo{volume}{84}}, \bibinfo{eid}{027302}
  (\bibinfo{year}{2011}), \eprint{1106.1528}.

\bibitem[{\citenamefont{Finkbeiner et~al.}(2012)\citenamefont{Finkbeiner,
  Galli, Lin, and Slatyer}}]{Finkbeiner:2011dx}
\bibinfo{author}{\bibfnamefont{D.~P.} \bibnamefont{Finkbeiner}},
  \bibinfo{author}{\bibfnamefont{S.}~\bibnamefont{Galli}},
  \bibinfo{author}{\bibfnamefont{T.}~\bibnamefont{Lin}}, \bibnamefont{and}
  \bibinfo{author}{\bibfnamefont{T.~R.} \bibnamefont{Slatyer}},
  \bibinfo{journal}{Phys.Rev.} \textbf{\bibinfo{volume}{D85}},
  \bibinfo{pages}{043522} (\bibinfo{year}{2012}), \eprint{1109.6322}.

\bibitem[{\citenamefont{{Slatyer}}(2013)}]{2013PhRvD..87l3513S}
\bibinfo{author}{\bibfnamefont{T.~R.} \bibnamefont{{Slatyer}}},
  \bibinfo{journal}{\prd} \textbf{\bibinfo{volume}{87}}, \bibinfo{eid}{123513}
  (\bibinfo{year}{2013}), \eprint{1211.0283}.

\bibitem[{\citenamefont{Galli et~al.}(2013)\citenamefont{Galli, Slatyer,
  Valdes, and Iocco}}]{Galli:2013dna}
\bibinfo{author}{\bibfnamefont{S.}~\bibnamefont{Galli}},
  \bibinfo{author}{\bibfnamefont{T.~R.} \bibnamefont{Slatyer}},
  \bibinfo{author}{\bibfnamefont{M.}~\bibnamefont{Valdes}}, \bibnamefont{and}
  \bibinfo{author}{\bibfnamefont{F.}~\bibnamefont{Iocco}},
  \bibinfo{journal}{Phys.Rev.} \textbf{\bibinfo{volume}{D88}},
  \bibinfo{pages}{063502} (\bibinfo{year}{2013}), \eprint{1306.0563}.

\bibitem[{\citenamefont{Lopez-Honorez et~al.}(2013)\citenamefont{Lopez-Honorez,
  Mena, Palomares-Ruiz, and Vincent}}]{Lopez-Honorez:2013cua}
\bibinfo{author}{\bibfnamefont{L.}~\bibnamefont{Lopez-Honorez}},
  \bibinfo{author}{\bibfnamefont{O.}~\bibnamefont{Mena}},
  \bibinfo{author}{\bibfnamefont{S.}~\bibnamefont{Palomares-Ruiz}},
  \bibnamefont{and} \bibinfo{author}{\bibfnamefont{A.~C.}
  \bibnamefont{Vincent}}, \bibinfo{journal}{JCAP}
  \textbf{\bibinfo{volume}{1307}}, \bibinfo{pages}{046} (\bibinfo{year}{2013}),
  \eprint{1303.5094}.

\bibitem[{\citenamefont{Madhavacheril et~al.}(2014)\citenamefont{Madhavacheril,
  Sehgal, and Slatyer}}]{Madhavacheril:2013cna}
\bibinfo{author}{\bibfnamefont{M.~S.} \bibnamefont{Madhavacheril}},
  \bibinfo{author}{\bibfnamefont{N.}~\bibnamefont{Sehgal}}, \bibnamefont{and}
  \bibinfo{author}{\bibfnamefont{T.~R.} \bibnamefont{Slatyer}},
  \bibinfo{journal}{Phys.Rev.} \textbf{\bibinfo{volume}{D89}},
  \bibinfo{pages}{103508} (\bibinfo{year}{2014}), \eprint{1310.3815}.

\bibitem[{\citenamefont{Valdes and Ferrara}(2008)}]{Valdes:2008cr}
\bibinfo{author}{\bibfnamefont{M.}~\bibnamefont{Valdes}} \bibnamefont{and}
  \bibinfo{author}{\bibfnamefont{A.}~\bibnamefont{Ferrara}},
  \bibinfo{journal}{Mon.Not.Roy.Astron.Soc.} \textbf{\bibinfo{volume}{387}},
  \bibinfo{pages}{8} (\bibinfo{year}{2008}), \eprint{0803.0370}.

\bibitem[{\citenamefont{Valdes et~al.}(2010)\citenamefont{Valdes, Evoli, and
  Ferrara}}]{Valdes:2009cq}
\bibinfo{author}{\bibfnamefont{M.}~\bibnamefont{Valdes}},
  \bibinfo{author}{\bibfnamefont{C.}~\bibnamefont{Evoli}}, \bibnamefont{and}
  \bibinfo{author}{\bibfnamefont{A.}~\bibnamefont{Ferrara}},
  \bibinfo{journal}{Mon.Not.Roy.Astron.Soc.} \textbf{\bibinfo{volume}{404}},
  \bibinfo{pages}{1569} (\bibinfo{year}{2010}), \eprint{0911.1125}.

\bibitem[{\citenamefont{{Furlanetto} and
  {Stoever}}(2010)}]{2010MNRAS.404.1869F}
\bibinfo{author}{\bibfnamefont{S.~R.} \bibnamefont{{Furlanetto}}}
  \bibnamefont{and} \bibinfo{author}{\bibfnamefont{S.~J.}
  \bibnamefont{{Stoever}}}, \bibinfo{journal}{\mnras}
  \textbf{\bibinfo{volume}{404}}, \bibinfo{pages}{1869} (\bibinfo{year}{2010}),
  \eprint{0910.4410}.

\bibitem[{\citenamefont{Evoli et~al.}(2012)\citenamefont{Evoli, Valdes,
  Ferrara, and Yoshida}}]{MNR:MNR20624}
\bibinfo{author}{\bibfnamefont{C.}~\bibnamefont{Evoli}},
  \bibinfo{author}{\bibfnamefont{M.}~\bibnamefont{Valdes}},
  \bibinfo{author}{\bibfnamefont{A.}~\bibnamefont{Ferrara}}, \bibnamefont{and}
  \bibinfo{author}{\bibfnamefont{N.}~\bibnamefont{Yoshida}},
  \bibinfo{journal}{Monthly Notices of the Royal Astronomical Society}
  \textbf{\bibinfo{volume}{422}}, \bibinfo{pages}{420} (\bibinfo{year}{2012}),
  ISSN \bibinfo{issn}{1365-2966},
  \urlprefix\url{http://dx.doi.org/10.1111/j.1365-2966.2012.20624.x}.

\bibitem[{\citenamefont{{Shull} and {van
  Steenberg}}(1985)}]{1985ApJ...298..268S}
\bibinfo{author}{\bibfnamefont{J.~M.} \bibnamefont{{Shull}}} \bibnamefont{and}
  \bibinfo{author}{\bibfnamefont{M.~E.} \bibnamefont{{van Steenberg}}},
  \bibinfo{journal}{\apj} \textbf{\bibinfo{volume}{298}}, \bibinfo{pages}{268}
  (\bibinfo{year}{1985}).

\bibitem[{\citenamefont{Slatyer}(2015)}]{inprep}
\bibinfo{author}{\bibfnamefont{T.~R.} \bibnamefont{Slatyer}},
  \bibinfo{journal}{to appear (Paper II)}  (\bibinfo{year}{2015}).

\bibitem[{\citenamefont{Cirelli et~al.}(2011)\citenamefont{Cirelli, Corcella,
  Hektor, Hutsi, Kadastik et~al.}}]{Cirelli:2010xx}
\bibinfo{author}{\bibfnamefont{M.}~\bibnamefont{Cirelli}},
  \bibinfo{author}{\bibfnamefont{G.}~\bibnamefont{Corcella}},
  \bibinfo{author}{\bibfnamefont{A.}~\bibnamefont{Hektor}},
  \bibinfo{author}{\bibfnamefont{G.}~\bibnamefont{Hutsi}},
  \bibinfo{author}{\bibfnamefont{M.}~\bibnamefont{Kadastik}},
  \bibnamefont{et~al.}, \bibinfo{journal}{JCAP}
  \textbf{\bibinfo{volume}{1103}}, \bibinfo{pages}{051} (\bibinfo{year}{2011}),
  \eprint{1012.4515}.

\bibitem[{\citenamefont{Zavala et~al.}(2010)\citenamefont{Zavala, Vogelsberger,
  and White}}]{Zavala:2009mi}
\bibinfo{author}{\bibfnamefont{J.}~\bibnamefont{Zavala}},
  \bibinfo{author}{\bibfnamefont{M.}~\bibnamefont{Vogelsberger}},
  \bibnamefont{and} \bibinfo{author}{\bibfnamefont{S.~D.~M.}
  \bibnamefont{White}}, \bibinfo{journal}{Phys. Rev.}
  \textbf{\bibinfo{volume}{D81}}, \bibinfo{pages}{083502}
  (\bibinfo{year}{2010}), \eprint{0910.5221}.

\bibitem[{\citenamefont{Hannestad and Tram}(2011)}]{Hannestad:2010zt}
\bibinfo{author}{\bibfnamefont{S.}~\bibnamefont{Hannestad}} \bibnamefont{and}
  \bibinfo{author}{\bibfnamefont{T.}~\bibnamefont{Tram}},
  \bibinfo{journal}{JCAP} \textbf{\bibinfo{volume}{1101}}, \bibinfo{pages}{016}
  (\bibinfo{year}{2011}), \eprint{1008.1511}.

\bibitem[{\citenamefont{{Chluba} and {Sunyaev}}(2012)}]{2012MNRAS.419.1294C}
\bibinfo{author}{\bibfnamefont{J.}~\bibnamefont{{Chluba}}} \bibnamefont{and}
  \bibinfo{author}{\bibfnamefont{R.~A.} \bibnamefont{{Sunyaev}}},
  \bibinfo{journal}{\mnras} \textbf{\bibinfo{volume}{419}},
  \bibinfo{pages}{1294} (\bibinfo{year}{2012}), \eprint{1109.6552}.

\bibitem[{\citenamefont{Chluba}(2013)}]{Chluba:2013wsa}
\bibinfo{author}{\bibfnamefont{J.}~\bibnamefont{Chluba}},
  \bibinfo{journal}{Mon.Not.Roy.Astron.Soc.} \textbf{\bibinfo{volume}{436}},
  \bibinfo{pages}{2232} (\bibinfo{year}{2013}), \eprint{1304.6121}.

\bibitem[{\citenamefont{Cirelli
  et~al.}(2009{\natexlab{a}})\citenamefont{Cirelli, Iocco, and
  Panci}}]{Cirelli:2009bb}
\bibinfo{author}{\bibfnamefont{M.}~\bibnamefont{Cirelli}},
  \bibinfo{author}{\bibfnamefont{F.}~\bibnamefont{Iocco}}, \bibnamefont{and}
  \bibinfo{author}{\bibfnamefont{P.}~\bibnamefont{Panci}},
  \bibinfo{journal}{JCAP} \textbf{\bibinfo{volume}{0910}}, \bibinfo{pages}{009}
  (\bibinfo{year}{2009}{\natexlab{a}}), \eprint{0907.0719}.

\bibitem[{\citenamefont{{Giesen} et~al.}(2012)\citenamefont{{Giesen},
  {Lesgourgues}, {Audren}, and {Ali-Ha{\"i}moud}}}]{2012JCAP...12..008G}
\bibinfo{author}{\bibfnamefont{G.}~\bibnamefont{{Giesen}}},
  \bibinfo{author}{\bibfnamefont{J.}~\bibnamefont{{Lesgourgues}}},
  \bibinfo{author}{\bibfnamefont{B.}~\bibnamefont{{Audren}}}, \bibnamefont{and}
  \bibinfo{author}{\bibfnamefont{Y.}~\bibnamefont{{Ali-Ha{\"i}moud}}},
  \bibinfo{journal}{JCAP} \textbf{\bibinfo{volume}{12}}, \bibinfo{eid}{008}
  (\bibinfo{year}{2012}), \eprint{1209.0247}.

\bibitem[{\citenamefont{{Farhang} et~al.}(2012)\citenamefont{{Farhang}, {Bond},
  and {Chluba}}}]{2012ApJ...752...88F}
\bibinfo{author}{\bibfnamefont{M.}~\bibnamefont{{Farhang}}},
  \bibinfo{author}{\bibfnamefont{J.~R.} \bibnamefont{{Bond}}},
  \bibnamefont{and} \bibinfo{author}{\bibfnamefont{J.}~\bibnamefont{{Chluba}}},
  \bibinfo{journal}{\apj} \textbf{\bibinfo{volume}{752}}, \bibinfo{eid}{88}
  (\bibinfo{year}{2012}), \eprint{1110.4608}.

\bibitem[{\citenamefont{Weniger et~al.}(2013)\citenamefont{Weniger, Serpico,
  Iocco, and Bertone}}]{Weniger:2013hja}
\bibinfo{author}{\bibfnamefont{C.}~\bibnamefont{Weniger}},
  \bibinfo{author}{\bibfnamefont{P.~D.} \bibnamefont{Serpico}},
  \bibinfo{author}{\bibfnamefont{F.}~\bibnamefont{Iocco}}, \bibnamefont{and}
  \bibinfo{author}{\bibfnamefont{G.}~\bibnamefont{Bertone}},
  \bibinfo{journal}{Phys.Rev.} \textbf{\bibinfo{volume}{D87}},
  \bibinfo{pages}{123008} (\bibinfo{year}{2013}), \eprint{1303.0942}.

\bibitem[{\citenamefont{Cirelli
  et~al.}(2009{\natexlab{b}})\citenamefont{Cirelli, Kadastik, Raidal, and
  Strumia}}]{Cirelli:2008pk}
\bibinfo{author}{\bibfnamefont{M.}~\bibnamefont{Cirelli}},
  \bibinfo{author}{\bibfnamefont{M.}~\bibnamefont{Kadastik}},
  \bibinfo{author}{\bibfnamefont{M.}~\bibnamefont{Raidal}}, \bibnamefont{and}
  \bibinfo{author}{\bibfnamefont{A.}~\bibnamefont{Strumia}},
  \bibinfo{journal}{Nucl. Phys.} \textbf{\bibinfo{volume}{B813}},
  \bibinfo{pages}{1} (\bibinfo{year}{2009}{\natexlab{b}}), \eprint{0809.2409}.

\bibitem[{\citenamefont{Boudaud et~al.}(2015)\citenamefont{Boudaud, Aupetit,
  Caroff, Putze, Belanger et~al.}}]{Boudaud:2014dta}
\bibinfo{author}{\bibfnamefont{M.}~\bibnamefont{Boudaud}},
  \bibinfo{author}{\bibfnamefont{S.}~\bibnamefont{Aupetit}},
  \bibinfo{author}{\bibfnamefont{S.}~\bibnamefont{Caroff}},
  \bibinfo{author}{\bibfnamefont{A.}~\bibnamefont{Putze}},
  \bibinfo{author}{\bibfnamefont{G.}~\bibnamefont{Belanger}},
  \bibnamefont{et~al.}, \bibinfo{journal}{Astron.Astrophys.}
  \textbf{\bibinfo{volume}{575}}, \bibinfo{pages}{A67} (\bibinfo{year}{2015}),
  \eprint{1410.3799}.

\bibitem[{\citenamefont{Lopez et~al.}(2015)\citenamefont{Lopez, Savage,
  Spolyar, and Adams}}]{Lopez:2015uma}
\bibinfo{author}{\bibfnamefont{A.}~\bibnamefont{Lopez}},
  \bibinfo{author}{\bibfnamefont{C.}~\bibnamefont{Savage}},
  \bibinfo{author}{\bibfnamefont{D.}~\bibnamefont{Spolyar}}, \bibnamefont{and}
  \bibinfo{author}{\bibfnamefont{D.~Q.} \bibnamefont{Adams}}
  (\bibinfo{year}{2015}), \eprint{1501.01618}.

\bibitem[{\citenamefont{Cline and Scott}(2013)}]{Cline:2013fm}
\bibinfo{author}{\bibfnamefont{J.~M.} \bibnamefont{Cline}} \bibnamefont{and}
  \bibinfo{author}{\bibfnamefont{P.}~\bibnamefont{Scott}},
  \bibinfo{journal}{JCAP} \textbf{\bibinfo{volume}{1303}}, \bibinfo{pages}{044}
  (\bibinfo{year}{2013}), \eprint{1301.5908}.

\bibitem[{\citenamefont{Ciafaloni et~al.}(2011)\citenamefont{Ciafaloni,
  Comelli, Riotto, Sala, Strumia et~al.}}]{Ciafaloni:2010ti}
\bibinfo{author}{\bibfnamefont{P.}~\bibnamefont{Ciafaloni}},
  \bibinfo{author}{\bibfnamefont{D.}~\bibnamefont{Comelli}},
  \bibinfo{author}{\bibfnamefont{A.}~\bibnamefont{Riotto}},
  \bibinfo{author}{\bibfnamefont{F.}~\bibnamefont{Sala}},
  \bibinfo{author}{\bibfnamefont{A.}~\bibnamefont{Strumia}},
  \bibnamefont{et~al.}, \bibinfo{journal}{JCAP}
  \textbf{\bibinfo{volume}{1103}}, \bibinfo{pages}{019} (\bibinfo{year}{2011}),
  \eprint{1009.0224}.

\bibitem[{\citenamefont{Aartsen et~al.}(2013)}]{Aartsen:2013dxa}
\bibinfo{author}{\bibfnamefont{M.}~\bibnamefont{Aartsen}} \bibnamefont{et~al.}
  (\bibinfo{collaboration}{IceCube}), \bibinfo{journal}{Phys.Rev.}
  \textbf{\bibinfo{volume}{D88}}, \bibinfo{pages}{122001}
  (\bibinfo{year}{2013}), \eprint{1307.3473}.

\bibitem[{\citenamefont{Aartsen et~al.}(2015{\natexlab{a}})}]{Aartsen:2015xej}
\bibinfo{author}{\bibfnamefont{M.}~\bibnamefont{Aartsen}} \bibnamefont{et~al.}
  (\bibinfo{collaboration}{IceCube}) (\bibinfo{year}{2015}{\natexlab{a}}),
  \eprint{1505.07259}.

\bibitem[{\citenamefont{Aartsen et~al.}(2015{\natexlab{b}})}]{Aartsen:2014hva}
\bibinfo{author}{\bibfnamefont{M.}~\bibnamefont{Aartsen}} \bibnamefont{et~al.}
  (\bibinfo{collaboration}{IceCube}), \bibinfo{journal}{Eur.Phys.J.}
  \textbf{\bibinfo{volume}{C75}}, \bibinfo{pages}{20}
  (\bibinfo{year}{2015}{\natexlab{b}}), \eprint{1406.6868}.

\bibitem[{\citenamefont{D'Agnolo and Ruderman}(2015)}]{D'Agnolo:2015koa}
\bibinfo{author}{\bibfnamefont{R.~T.} \bibnamefont{D'Agnolo}} \bibnamefont{and}
  \bibinfo{author}{\bibfnamefont{J.~T.} \bibnamefont{Ruderman}}
  (\bibinfo{year}{2015}), \eprint{1505.07107}.

\bibitem[{\citenamefont{Izaguirre et~al.}(2015)\citenamefont{Izaguirre,
  Krnjaic, Schuster, and Toro}}]{Izaguirre:2015yja}
\bibinfo{author}{\bibfnamefont{E.}~\bibnamefont{Izaguirre}},
  \bibinfo{author}{\bibfnamefont{G.}~\bibnamefont{Krnjaic}},
  \bibinfo{author}{\bibfnamefont{P.}~\bibnamefont{Schuster}}, \bibnamefont{and}
  \bibinfo{author}{\bibfnamefont{N.}~\bibnamefont{Toro}}
  (\bibinfo{year}{2015}), \eprint{1505.00011}.

\bibitem[{\citenamefont{Adriani et~al.}(2009)}]{Adriani:2008zr}
\bibinfo{author}{\bibfnamefont{O.}~\bibnamefont{Adriani}} \bibnamefont{et~al.}
  (\bibinfo{collaboration}{PAMELA}), \bibinfo{journal}{Nature}
  \textbf{\bibinfo{volume}{458}}, \bibinfo{pages}{607} (\bibinfo{year}{2009}),
  \eprint{0810.4995}.

\bibitem[{\citenamefont{{Ackermann} et~al.}(2012)\citenamefont{{Ackermann},
  {Ajello}, {Allafort}, {Atwood}, {Baldini}, {Barbiellini}, {Bastieri},
  {Bechtol}, {Bellazzini}, {Berenji} et~al.}}]{2012PhRvL.108a1103A}
\bibinfo{author}{\bibfnamefont{M.}~\bibnamefont{{Ackermann}}},
  \bibinfo{author}{\bibfnamefont{M.}~\bibnamefont{{Ajello}}},
  \bibinfo{author}{\bibfnamefont{A.}~\bibnamefont{{Allafort}}},
  \bibinfo{author}{\bibfnamefont{W.~B.} \bibnamefont{{Atwood}}},
  \bibinfo{author}{\bibfnamefont{L.}~\bibnamefont{{Baldini}}},
  \bibinfo{author}{\bibfnamefont{G.}~\bibnamefont{{Barbiellini}}},
  \bibinfo{author}{\bibfnamefont{D.}~\bibnamefont{{Bastieri}}},
  \bibinfo{author}{\bibfnamefont{K.}~\bibnamefont{{Bechtol}}},
  \bibinfo{author}{\bibfnamefont{R.}~\bibnamefont{{Bellazzini}}},
  \bibinfo{author}{\bibfnamefont{B.}~\bibnamefont{{Berenji}}},
  \bibnamefont{et~al.}, \bibinfo{journal}{Physical Review Letters}
  \textbf{\bibinfo{volume}{108}}, \bibinfo{eid}{011103} (\bibinfo{year}{2012}),
  \eprint{1109.0521}.

\bibitem[{\citenamefont{Accardo et~al.}(2014)\citenamefont{Accardo, Aguilar,
  Aisa, Alpat, Alvino, Ambrosi, Andeen, Arruda, Attig, Azzarello
  et~al.}}]{PhysRevLett.113.121101}
\bibinfo{author}{\bibfnamefont{L.}~\bibnamefont{Accardo}},
  \bibinfo{author}{\bibfnamefont{M.}~\bibnamefont{Aguilar}},
  \bibinfo{author}{\bibfnamefont{D.}~\bibnamefont{Aisa}},
  \bibinfo{author}{\bibfnamefont{B.}~\bibnamefont{Alpat}},
  \bibinfo{author}{\bibfnamefont{A.}~\bibnamefont{Alvino}},
  \bibinfo{author}{\bibfnamefont{G.}~\bibnamefont{Ambrosi}},
  \bibinfo{author}{\bibfnamefont{K.}~\bibnamefont{Andeen}},
  \bibinfo{author}{\bibfnamefont{L.}~\bibnamefont{Arruda}},
  \bibinfo{author}{\bibfnamefont{N.}~\bibnamefont{Attig}},
  \bibinfo{author}{\bibfnamefont{P.}~\bibnamefont{Azzarello}},
  \bibnamefont{et~al.} (\bibinfo{collaboration}{(AMS Collaboration)}),
  \bibinfo{journal}{Phys. Rev. Lett.} \textbf{\bibinfo{volume}{113}},
  \bibinfo{pages}{121101} (\bibinfo{year}{2014}),
  \urlprefix\url{http://link.aps.org/doi/10.1103/PhysRevLett.113.121101}.

\bibitem[{\citenamefont{Cholis and Hooper}(2013)}]{Cholis:2013psa}
\bibinfo{author}{\bibfnamefont{I.}~\bibnamefont{Cholis}} \bibnamefont{and}
  \bibinfo{author}{\bibfnamefont{D.}~\bibnamefont{Hooper}},
  \bibinfo{journal}{Phys.Rev.} \textbf{\bibinfo{volume}{D88}},
  \bibinfo{pages}{023013} (\bibinfo{year}{2013}), \eprint{1304.1840}.

\bibitem[{\citenamefont{{Seager} et~al.}(1999)\citenamefont{{Seager},
  {Sasselov}, and {Scott}}}]{Seager:1999}
\bibinfo{author}{\bibfnamefont{S.}~\bibnamefont{{Seager}}},
  \bibinfo{author}{\bibfnamefont{D.~D.} \bibnamefont{{Sasselov}}},
  \bibnamefont{and} \bibinfo{author}{\bibfnamefont{D.}~\bibnamefont{{Scott}}},
  \bibinfo{journal}{\apjl} \textbf{\bibinfo{volume}{523}}, \bibinfo{pages}{L1}
  (\bibinfo{year}{1999}), \eprint{arXiv:astro-ph/9909275}.

\bibitem[{\citenamefont{{Chluba} and {Thomas}}(2011)}]{Chluba:2010ca}
\bibinfo{author}{\bibfnamefont{J.}~\bibnamefont{{Chluba}}} \bibnamefont{and}
  \bibinfo{author}{\bibfnamefont{R.~M.} \bibnamefont{{Thomas}}},
  \bibinfo{journal}{\mnras} \textbf{\bibinfo{volume}{412}},
  \bibinfo{pages}{748} (\bibinfo{year}{2011}), \eprint{1010.3631}.

\bibitem[{\citenamefont{Ali-Haimoud and Hirata}(2011)}]{AliHaimoud:2010dx}
\bibinfo{author}{\bibfnamefont{Y.}~\bibnamefont{Ali-Haimoud}} \bibnamefont{and}
  \bibinfo{author}{\bibfnamefont{C.~M.} \bibnamefont{Hirata}},
  \bibinfo{journal}{Phys. Rev.} \textbf{\bibinfo{volume}{D83}},
  \bibinfo{pages}{043513} (\bibinfo{year}{2011}), \eprint{1011.3758}.

\bibitem[{\citenamefont{{Lewis} et~al.}(2000)\citenamefont{{Lewis},
  {Challinor}, and {Lasenby}}}]{2000ApJ...538..473L}
\bibinfo{author}{\bibfnamefont{A.}~\bibnamefont{{Lewis}}},
  \bibinfo{author}{\bibfnamefont{A.}~\bibnamefont{{Challinor}}},
  \bibnamefont{and}
  \bibinfo{author}{\bibfnamefont{A.}~\bibnamefont{{Lasenby}}},
  \bibinfo{journal}{\apj} \textbf{\bibinfo{volume}{538}}, \bibinfo{pages}{473}
  (\bibinfo{year}{2000}), \eprint{arXiv:astro-ph/9911177}.

\end{thebibliography}

\end{document}